\documentclass[a4paper,11pt]{article}

\usepackage{feynmp}
\usepackage{graphicx}
\usepackage{fullpage}
\usepackage{framed}

\usepackage{mystyle}
\graphicspath{{./figures/}}

\usepackage[table]{xcolor}
\usepackage{booktabs}
\usepackage{multirow}

\usepackage{pdflscape}

\usepackage{tikz}
\usetikzlibrary{decorations.pathmorphing, decorations.markings, calc, fadings,
decorations.pathreplacing, patterns,shapes.arrows,positioning,3d,arrows,decorations.markings,shadings,backgrounds}

\usepackage[displaymath, mathlines]{lineno}
\usepackage{tensor}
\usepackage{cancel}

\usepackage{framed}
\usepackage{url}
\usepackage{caption}
\usepackage[justification=centering]{subcaption}

\title{\sffamily Effective field theory, electric dipole moments and electroweak baryogenesis}
\author{Csaba Balazs$^a$, Graham White$^a$ and  Jason Yue$^{b,c}$}
\affiliation[]{
$^a$ARC Centre of Excellence for Particle Physics at the Terascale
School of Physics and Astronomy, Monash University
Vicotria 3800, Australia\\
$^b$Department of Physics, National Taiwan Normal University, Taipei 116, Taiwan\\
$^c$ARC Centre of Excellence for Particle Physics at the Terascale, School of Physics, The University of Sydney, NSW 2006, Australia
}

\emailAdd{csaba.balazs@monash.edu}
\emailAdd{graham.white@monash.edu}
\emailAdd{jason.yue@ntnu.edu.tw}

\abstract{
Negative searches for permanent electric dipole moments (EDMs) heavily constrain models of baryogenesis utilising various higher dimensional charge and parity violating (CPV) operators. 
Using effective field theory, we create a model independent connection between these EDM constraints and the baryon asymmetry of the universe (BAU) produced during a strongly first order electroweak phase transition. 
The thermal aspects of the high scale physics driving the phase transition are paramaterised by the usual kink solution for the bubble wall profile.
We find that operators involving derivatives of the Higgs field yield CPV contributions to the BAU containing derivatives of the Higgs vacuum expectation value (vev), while non-derivative operators lack such contributions.  Consequently, derivative operators cannot be eliminated in terms of non-derivative operators (via the equations of motion) if one is agnostic to the new physics that leads to the phase transition. 
Thus, we re-classify the independent dimension six operators, restricting ourselves to third generation quarks, gauge bosons and the Higgs.
Finally, we calculate the BAU (as a function of the bubble wall width and the cutoff) for a derivative and a non-derivative operator, and relate it to the EDM constraints. 
}

\begin{document}

\maketitle


\section{Introduction}\label{sec:intro}

  



The Higgs boson  discovered at the Large Hadron Collider (LHC) \cite{Aad:2012tfa, Chatrchyan:2012xdj} closely resembles that of the Standard Model (SM).  This rules out the mechanism of electroweak baryogenesis (cf. \cite{White:2016nbo} and references therein for a pedagogical review) within the SM because with a Higgs mass of 125 GeV \cite{Aad:2015zhl} the electroweak phase transition (EWPT) does not provide a sufficient departure from equilibrium \cite{Rummukainen:1998as}.  The SM also falls short in the amount of charge (C) and charge-parity (CP) violation to generate the observed baryon asymmetry of the universe (BAU) \cite{Gavela:1993ts,Konstandin:2003dx}.  These two facts alone are enough to motivate the existence of new physics responsible for baryon asymmetry.  

Physics models entailing new particles or interactions can introduce charge-parity violating (CPV) phases to assist explaining the observed BAU \cite{Ade:2015xua} via electroweak baryogenesis. 
The use of effective field theories (EFTs) allows one to test a large class of models without adhering to a specific model or framework. This greatly facilitates the connection with experimental constraints.
%
Under this motivation, we consider an extension of the Standard Model by effective dimension six operators.  To achieve electroweak baryogenesis, one typically utilises two such higher dimensional operators\footnote{For an approach where EWBG is achieved without adding particle content to the SM nor invoking higher dimensional operators, see \cite{Kobakhidze:2015xlz}.} to simultaneously generate enough $CP$ violation and a strongly first order phase transition (SFOPT) at the electroweak scale (cf. \cite{Bodeker:2004ws,Fromme:2006wx,Huang:2015izx} and references therein).  Considerable amount of literature have been devoted to generate sufficient CPV via dimension six operators \cite{Engel:2013lsa, Chupp:2014gka, Chien:2015xha, Brod:2013cka}, whilst evading ever tighter constraints from searches for permanent electric dipole moments (EDMs).  Similarly, studies of a SFOPT catalysed by dimension six operators \cite{Noble:2007kk, Delaunay:2007wb} (particularly applied to top-Higgs sector \cite{Cirigliano:2016njn, Cirigliano:2016nyn}) place a bound of $\Lambda\lesssim 800$ GeV on the scale of new physics that could boost the strength of the phase transition \cite{Grojean:2004xa}.

In this work, we argue that it possible to build a relatively direct bridge between the EDM constraints on a higher dimensional operator and the maximal baryon asymmetry produced by such an operator by assuming a strongly first order phase transition (which is parametrised by the bubble wall width, velocity, etc.)  \cite{Fuyuto:2015ida, Huber:2006ri, Shu:2013uua,Zhang:1993vh}. This bridge  can be used to then classify the UV completion(s) corresponding to the EFT (for some examples see \cite{Huang:2015bta, Damgaard:2015con, Bernal:2012gv, Zhang:2012cd})\footnote{See \cite{Herrero-Garcia:2016uab} for an approach of connecting EFTs and UV complete models in the context of Higgs flavour violation.}. 
%
While building the above bridge, we point out that the degeneracy between certain higher dimensional operators is lifted.  Usually, derivative operators are traded to non-derivative ones via the classical equations of motion. However, such degeneracy may be broken in BAU calculations since the CPV sources corresponding to these operators have different dependencies on the assumed profile of the space-time varying vacuum.
%
It is necessary then to extend the higher dimensional $CP$ violating operator basis (cf. e.g. \cite{Yang:1997iv,Whisnant:1997qu,AguilarSaavedra:2009mx}) that is capable of generating the baryon asymmetry. 

The most promising operators for BAU generation are those that contain at least one Higgs field to accommodate $CP$ violating interactions with a space-time varying bubble wall as well as a strongly coupled SM field, i.e. a top quark or a gauge boson. The resonant enhancement of such interaction during the electroweak phase transition becomes the most efficient mode for baryogensis. Consequently, two qualitatively different operators are chosen within the new catalogue of the operators presented in this paper to demonstrate the aforementioned bridge. 
One of the operators chosen is normally considered redundant due to the equations of motion. It involves a derivative coupling 
to the Higgs and the result is an increased sensitivity to the width of the electroweak bubble wall.
The respective baryon asymmetries are calculated show that current EDM measurements can meaningfully constrain the available parameter space.

The structure of this paper is outlined as follows. In Sec.~\ref{sec:lift_degen} we demonstrate that the redundancy between various operators is lost during the electroweak phase transition. We then catalogue the full set of $CP$ violating dimension six operators that are candidates for producing the BAU via the electroweak mechanism in Sec.~\ref{sec:op}. The  $CP$ violating sources are calculated using the closed time path formalism in Sec.~\ref{sec:ewbg_eft}, with their respective EDM constraints derived subsequently in Sec.~\ref{sec:edms}. We present resulting BAU in Sec.~\ref{sec:results} before briefly discussing the possibility of space-time varying masses of heavy particles in Sec.~\ref{sec:spacetime_dep}. Finally we conclude with Sec.~\ref{sec:concl}.


\section{Removing redundancies of operators with derivative coupling to the Higgs}\label{sec:lift_degen}
A successful explanation of the BAU necessarily fulfils the three Sakharov conditions \cite{Sakharov:1967dj}:
\begin{enumerate}
    \item[(1)] baryon number violation,
    \item[(2)] charge and charge-parity violation, and
    \item[(3)] departure from thermal equilibrium.
\end{enumerate}
In electroweak baryogenesis, the $SU(2)$ sphalerons are responsible for meeting the first condition as the anomalous baryon number violating processes become unsuppressed at high temperature. In the SM, the second condition is met through a $CP$ violating phase in the Cabibbo–Kobayashi–Maskawa (CKM) matrix, but it is too feeble to provide enough baryon asymmetry. The third condition also fails in the SM as the Higgs mass is too heavy to catalyse a strongly first order electroweak phase transition. 

The second and third conditions can be satisfied within the SMEFT framework by adding higher dimensional operators to the SM Lagrangian
\begin{equation}
  \mathcal{L} = \mathcal{ L}_{\text{SM}} + \frac{c_{CPV} }{\Lambda _{CPV}^2 } \mathcal{ O}_{D=6,\text{CPV}} + \sum _{n,m\in \mathbb{N}} \frac{c_{n,m}}{\Lambda ^n _m} \mathcal{O}_{\Delta V, D=4+n}^{(m)} ,
\end{equation}
where $\mathcal{O}_{D=6,\text{CPV}}$ is an operator\footnote{We note that one can in principle have many such operators. The approach we adopt here is to inspect each one separately as a sole source of CPV (in addition to the CKM phase).} contributing to both the BAU as well as EDMs, and the set of operators $\mathcal{O}^{(m)}_{\Delta V,D=4+n}$ ensure that the EWPT is strongly first order (see for example \cite{Bodeker:2004ws,Fromme:2006wx}).
In general, the $\mathcal{O}_{D=6,\text{CPV}}$ operator may contain derivatives, and there can be a single or several $\mathcal{O}^{(m)}_{\Delta V,D=4+n}$ operators, each possibly with a different cutoff scales. 

Usually, the classical equations of motion are used to eliminate derivative operators as redundant.  However, we will show in this section qualitatively (numerically in a subsequent section) that one should exercise caution when eliminating derivative operators with EOMs for baryon asymmetry calculations within the EFT framework.
The reason for this is because the Sakharov conditions are met only if {\it both} the operators $\mathcal{O}^{(m)}_{\Delta V,D=4+n}$ and $\mathcal{O}_{D=6,\text{CPV}}$ exist.  
%
%
For a concrete example consider an example operator of the class $\mathcal{O}_{D=6,\text{CPV}}$
\begin{equation}
    \mathcal{O}_{DD} = \overline{Q}_Lt_R D_\mu D^\mu H \ .
\end{equation}
The derivatives on the Higgs in the above operator can be typically eliminated by making use of the field equations
\begin{equation}
  D_\mu D^\mu H = - \frac{\partial \mathcal{L}_{\text{SM}}}{ \partial H^\dagger } + \mathcal{O}\left( \frac{1}{\Lambda} \right) . \label{fieldeqns}
\end{equation}
Explicitly, one can use the classical equations of motion to rewrite $\mathcal{O}_{DD}$ as
\begin{eqnarray}
  \frac{1}{\Lambda ^2}\mathcal{O}_{DD} &=& \frac{1}{\Lambda ^2} \overline{Q}_L t_R D_\mu D^\mu H \nonumber \\ &\rightarrow & \frac{1}{\Lambda ^2} \overline{Q}_L t_R \left(\frac{\partial \mathcal{L}_{\text{SM}}}{\partial H^\dagger} + \frac{\partial }{\partial H^\dagger} \sum _{n,m} \frac{c_{n,m}}{\Lambda ^n _m}  \mathcal{O}_{\Delta V, D=4+n}^{(m)} + \cdots  \right) \label{eq:redundancy} \ .
\end{eqnarray}
After substituting in the SM Lagrangian and evaluating the derivatives, the operator $\mathcal{O}_{DD}$ reads 
\begin{equation}    \label{eq:ODD_EOM}
  \begin{aligned}
    \mathcal{O}_{DD} =&~ \overline{Q}_L H^\dagger t_R \left(\mu^2-\lambda H^\dagger H\right) -\left(\overline{Q}_{L}  t_R\right)_i \epsilon^{ij} \left(\overline{L}Y_ee_R\right)_j \\
    -&~ \left(\overline{Q}_{L}  t_R\right)_i \epsilon^{ij}\left(\overline{u}_R Y_u^\dagger Q\right)_j -  \left(\overline{Q}_{L}  t_R\right)_i \epsilon^{ij}\left(\overline{Q}_i Y_d^\dagger d_R\right)_j + { \mathcal{O}\left(\frac{1}{\Lambda^2} \right)  } \ . 
  \end{aligned} 
\end{equation}
In EDM calculations, operators suppressed by $\Lambda^{-n}$ can be safely neglected in Eq. \ref{eq:redundancy}. The use of the equations of motion to relate different operators is justified in this context since a hierarchy of scale is established between the constant Higgs vacuum expectation value (vev), $ \langle H(x) \rangle = v$, and the cutoff $\Lambda$.

In BAU calculations, however, the $CP$ violating sources typically depend on the derivative of the space-time varying vacuum during the phase transition.  For example, for the supposedly degenerate operators shown in Eq. \ref{eq:redundancy}, one can derive the $CP$ violating sources to lowest order in the inverse cutoff \footnote{The reader is referred to (\ref{eq:def_I}) for the explicit form of the sources.}:
\begin{eqnarray}\label{eq:source_cpv}
S_{\mathcal{O}_{DD}} ^{\cancel{CP}} 
&\sim&  \frac{1}{\Lambda ^2} \left[v(x) \partial _t \left(\partial _\mu \partial ^\mu v(x)  \right)-\partial _t v(x) \left(\partial _\mu \partial ^\mu v(x)  \right)\right] , \nonumber \\
S_{\mathcal{O}_{\partial V/ \partial H}} ^{\cancel{CP}} 
&\sim&  \frac{1}{\Lambda ^2}\left[v(x) \partial _t \left(\left. \frac{\partial V_{\rm SM}}{\partial H}  \right|_{v(x)}\right) -\partial _t v(x)   \left(\left. \frac{\partial V_{\rm SM}}{\partial H}  \right|_{v(x)}\right) \right] + {\cal O} \left( \frac{1}{\Lambda ^4} \right) . \label{eq:approxCPVsources}
\end{eqnarray}
One can immediately see that the two expressions do not agree in general. This is made explicit when a specific form for the Higgs profile of  $v(x)$ is introduced to describe bubble formation during the electroweak phase transition.  The profile is
a is a stationary field configuration in the finite temperature effective action which interpolates the false vacuum to the true vacuum. Assuming an $O(3)$ symmetry, the \textit{bounce solution} takes the form:
\begin{equation}
  v(z) \approx \frac{v(T)}{2} \left[ 1+\tanh \left(  \frac{z}{L_w}\right)  \right] \ . \label{eq:kinkapprox}
\end{equation}
Here, $L_w$ measures the width of the bubble wall and $z$ parametrises the distance perpendicular to the wall. With both $L_w$ and $v(T)$  determined by the operators $ \mathcal{O}_{\Delta V, D=4+n}^{(m)}$, there are no free parameters left in Eq.~\ref{eq:kinkapprox}.  
Using Eq.~\ref{eq:kinkapprox} in Eq.~\ref{eq:approxCPVsources} does not yield the same result, not even approximately. This is due to the fact that $S_{\mathcal{O}_{DD}} ^{\cancel{CP}}$ and $S_{\mathcal{O}_{\partial V/ \partial H}} ^{\cancel{CP}}$ have different dependencies on $L_w$ and $v(T)$. Specifically, the CPV source resulting from the operator with derivative coupling to the Higgs, ${\cal O}_{DD}$, has a cubic sensitivity to the bubble wall width whereas the non-derivative operator has a quartic sensitivity to the value of $v(T)$. The strength of the CPV source due to the CPV operators then become very sensitive to the exact structure of the set of operators $ \mathcal{O}_{\Delta V, D=4+n}^{(m)}$ rather than the ${\cal O}(\Lambda ^{-4})$ sensitivity that occurs in EDM calculations.

For each $\mathcal{O}_{D=6,\text{CPV}}$, one could in principle consider every single possibility for the set $ \mathcal{O}_{\Delta V, D=4+n}^{(m)}$ and their Wilson coefficients to calculate $L_w$ and $v(T)$. However, by ignoring the precise structure of $ \mathcal{O}_{\Delta V, D=4+n}^{(m)}$ and instead leaving $L_w$ and $v(T)$ as free parameters, we can then draw as direct a bridge as possible between BAU calculations and EDM limit. The result is that redundancy between derivative and non-derivative operators is lifted. This is due to the sensitivity of $CP$ violating sources to $ \mathcal{O}_{\Delta V, D=4+n}^{(m)}$ being much sharper than the expected $\Lambda ^{-4}$ sensitivity. This behaviour is attributed to the derivative structure of the operators and the changing the profile of the vev during the EWPT as controlled by $L_w$ and $v(T)$.

\section{Operators classification}\label{sec:op}


In this section, we classify and count the operators involving derivative couplings with Higgs which can no longer be considered redundant.  In electroweak barogenesis, the SM fields whose contribution to the BAU are suppressed by small Yukawa couplings can be neglected. This means that only the left handed third generation quark doublet, the right-handed top, the Higgs and gauge bosons need be considered. With the symbolic meanings of $\psi$, $D$, $F$ and $H$ applied to fermions, derivative operators, (dual) field strength tensors and Higgs operators respectively, one should obtain 12 operator classes, 8 of which involve the Higgs
\begin{equation}
  \begin{aligned}
    & H^6,\quad  H^4D^2,\quad H^2 D^4, \quad FH^2D^2 , \quad  \psi^2 H^3,\quad F^2 H^2,\quad \psi^2 H^2 D,\quad \psi^2 H D^2,\quad \psi^2 HF,\\ & F^2D^2, \quad \psi^4, \quad
    \psi^2 D F, \quad  F^3 .
  \end{aligned}
\end{equation}
In order for the contributions to the BAU be resonantly enhanced, a $CP$ violating operator must involve at least one space-time varying Higgs operator and one other field.  
Terms with single $H$ cannot appear without a $\psi^2$ combination to cancel the $SU(2)$ charge.  Therefore, terms such as $FD^2 H$ should not appear.  Under these constraints the possible classes of operators are
\begin{equation}
  \begin{aligned}
H^4D^2,\quad H^2 D^4, \quad \psi ^2 H^3 ,  \quad FH^2D^2 , \quad F^2 H^2,\quad \psi^2 H^2 D,\quad \psi^2 H D^2,\quad \psi^2 HF .
  \end{aligned}
\end{equation}
We take the $CP$-odd operators from \cite{Yang:1997iv}, while the $CP$-even analogue is given in \cite{Whisnant:1997qu} (see also  \cite{Nomura:2009tw,AguilarSaavedra:2009mx}.)   
We  list in Tab.~\ref{tab:dim_six_op} the operators satisfying the above constraints.  We find 34 in total that fulfil all of our constraints, including 19 with higher derivative couplings that are usually considered redundant. We considered operators with $ D_\mu D^2$ and not $D_\mu D^2$ as the latter can be formed by taking the sum of the first operator and an operator involving the field strength tensor.  We will select two qualitatively different operators for an extensive study ---  one with a second derivative coupling, $\mathcal{ O}_{DD}$, and one with no derivative couplings $\mathcal{ O}_{t1}$. 

\begin{table}[!h]
\begin{center}\renewcommand{\arraystretch}{1.3}
\begin{tabular}[c]{|c|c| c|c|}
\hline
\multicolumn{2}{|c|}{$ H^2D^4$}
&\multicolumn{2}{c|}{$H^4 D^2$}
\\ \hline
\rowcolor{gray!20}
$\mathcal{O}_{H^2D^4}^{(1)}$       & $\left(D^4 H^\dagger\right)  H$
&$\mathcal{O}_{H^4D^2}^{(1a)}$      &  $ (H^\dagger H) H^\dagger D^2 H $
\\ \hline
$\mathcal{O}_{H^2D^4}^{(2)}$       & $\left(D^2 D_\mu H^\dagger\right) D^\mu  H$
&$\mathcal{O}_{H^4D^2}^{(2a)}$      &  $ (H^\dagger H) D^\mu H^\dagger D_\mu H $
\\ \hline
\rowcolor{gray!20}
$\mathcal{O}_{H^2D^4}^{(3a)}$       & $(D^2 H^\dagger)( D^2 H)$
&$\mathcal{O}_{H^4D^2}^{(2b)}$     &$ (H^\dagger \overset{\leftrightarrow}{ D^\mu } H)  (H^\dagger \overset{\leftrightarrow} D_\mu  H )$
\\\hline
$\mathcal{O}_{H^2D^4}^{(3b)}$       & $(D^\mu D^\nu H^\dagger)( D_\mu D_\nu H)$ 
&\multicolumn{2}{c|}{}
\\\hline

\end{tabular}\\[1cm]

\fontsize{10}{11}
\renewcommand{\arraystretch}{1.3}
\begin{tabular}[c]{|c|c| c|c| c|c|}
\hline
\multicolumn{2}{|c|}{$\psi^2 H^3$}
&\multicolumn{2}{c|}{$\psi^2 H^2D$}
&\multicolumn{2}{c|}{$\psi^2 HD^2$}
\\ \hline
\rowcolor{gray!20}
$\mathcal{O}_{t1}$       & $\left( H^\dagger H \right) \left(\overline{Q}_L \tilde{H}t_R  \right)$
&$\mathcal{O}_{Hq}^{(1)}$   &  $\left( H^\dagger i \overset{\leftrightarrow} D_\mu H  \right) \left(\overline{Q}_L \gamma^\mu Q_L\right)$
&$\mathcal{O}_{\sigma DD}$           & $\left( \overline{Q}_L \sigma^{\mu\nu}     t_R \right) D_\mu D_\nu \tilde{H}  $

\\ \hline
\multicolumn{2}{|c|}{}
&$\mathcal{O}_{Hq}^{(3)}$   &  $\left( H^\dagger i \overset{\leftrightarrow} {D_\mu^i} H  \right) \left(\overline{Q}_L \gamma^\mu \tau^i Q_L\right)$
&$\mathcal{O}_{\sigma DD}$  & $\left( \overline{Q}_L \sigma^{\mu\nu} \overset{\leftrightarrow} D_\mu    t_R \right)  D_\nu \tilde{H}  $
\\ \hline
\rowcolor{gray!20}
\multicolumn{2}{|c|}{}
&$\mathcal{O}_{Ht}$         &  $\left( H^\dagger i \overset{\leftrightarrow} D_\mu H  \right) \left(\overline{t}_R \gamma^\mu t_R\right)$
&$\mathcal{O}_{DD}$         & $\left( \overline{Q}_L      t_R \right)  D^\mu D_\mu \tilde{H}  $
\\ \hline
\multicolumn{2}{|c|}{}
&\multicolumn{2}{c|}{}
&$\mathcal{O}_{DtDH}$           & $\left( \overline{Q}_L  \overset{\leftrightarrow} D_\mu    t_R \right)  D^\mu \tilde{H}  $\\\hline

\end{tabular}\\[1cm]
\begin{tabular}[c]{|c| c| c|  c| c | c |c }
\hline
\multicolumn{2}{|c|}{$F^2 H^2$}
&\multicolumn{2}{c|}{$\psi^2 HF$}
&\multicolumn{2}{c|}{$F H^2D^2$}
\\ \hline
\rowcolor{gray!20}
$\mathcal{O}_{HG}$          & $\left( H^\dagger H \right) G^a_{\mu\nu} G^{a\mu\nu}$
&$\mathcal{O}_{tG}$         & $\left( \overline{Q}_L \sigma^{\mu\nu} T^a t_R \right) \tilde{H} G^a_{\mu\nu} $
&$\mathcal{O}_{D^2 HW}^{(1)}$   &  $W_{\mu\nu}^i (D^\mu H^\dagger) \tau^i (D^\nu H )$
\\ \hline
$\mathcal{O}_{H\tilde{G}}$  & $\left( H^\dagger H \right) G^a_{\mu\nu} \tilde{G}^{a\mu\nu}$
&$\mathcal{O}_{tW}$         & $\left( \overline{Q}_L \sigma^{\mu\nu} \tau^i t_R \right) \tilde{H} W^i_{\mu\nu} $
&$\mathcal{O}_{D^2 HW}^{(2)}$   & $D^\mu W_{\mu\nu}^i ( H^\dagger  \overset{\leftrightarrow}{D^{i,\nu} } H )$
\\ \hline
\rowcolor{gray!20}
$\mathcal{O}_{HW}$          & $\left( H^\dagger H \right) W^i_{\mu\nu} W^{i\mu\nu}$
&$\mathcal{O}_{tB}$         & $\left( \overline{Q}_L \sigma^{\mu\nu}  t_R \right) \tilde{H} B_{\mu\nu} $
&$\mathcal{O}_{D^2 H\tilde{W}}^{(1)}$   & $\tilde{W}_{\mu\nu}^i (D^\mu H^\dagger) \tau^i (D^\nu H) $ 
\\ \hline
$\mathcal{O}_{H\tilde{W}}$  & $\left( H^\dagger H \right) W^i_{\mu\nu} \tilde{W}^{i\mu\nu}$
&\multicolumn{2}{c|}{} 
&$\mathcal{O}_{D^2 H\tilde{W}}^{(2)}$   & $D^\mu \tilde{W}_{\mu\nu}^i ( H^\dagger \overset{\leftrightarrow}{D^{i,\nu}}  H) $  
\\ \hline
\rowcolor{gray!20}
$\mathcal{O}_{HB}$          & $\left( H^\dagger H \right) B_{\mu\nu} B^{\mu\nu}$
&\multicolumn{2}{c|}{}
&$\mathcal{O}_{D^2 BH}^{(1)}$  & $B_{\mu\nu} (D^\mu H^\dagger)  (D^\nu H) $  
\\ \hline
$\mathcal{O}_{H\tilde{B}}$  & $\left( H^\dagger H \right) B_{\mu\nu} \tilde{B}^{\mu\nu}$
&\multicolumn{2}{c|}{}
&$\mathcal{O}_{D^2 BH}^{(2)}$  & $D^\mu B_{\mu\nu} ( H^\dagger \overset{\leftrightarrow}{D^\nu}  H) $   
\\ \hline
\rowcolor{gray!20}
$\mathcal{O}_{HWB}$         & $\left( H^\dagger \tau^i H \right) W^i_{\mu\nu} B^{\mu\nu}$
&\multicolumn{2}{c|}{}
&$\mathcal{O}_{D^2 \tilde{B}H}^{(1)}$  &$\tilde{B}_{\mu\nu} (D^\mu H^\dagger)(  D^\nu H )$
\\ \hline
$\mathcal{O}_{H\tilde{W}B}$ & $\left( H^\dagger \tau^i H \right) \tilde{W}^i_{\mu\nu} B^{\mu\nu}$
&\multicolumn{2}{c|}{}
&$\mathcal{O}_{D^2 \tilde{B}H}^{(2)}$  &$D^\mu \tilde{B}_{\mu\nu} ( H^\dagger \overset{\leftrightarrow}{D^\nu} H )$
\\ \hline
\end{tabular}
\caption{
List of dimension six operators based on \cite{Grzadkowski:2010es} involving at least one Higgs and one other field that is either a Standard Model gauge boson or a top quark.  Since redundancies due to the equations of motion are no long applicable, one has to be cautious with the classes involving (i)  two derivatives acting a  Higgs field and (ii) one derivative acting on either gauge field or fermion field. Here we follow the definition that $H^\dagger i  \overset{\leftrightarrow}{D_\mu} H := i H ^\dagger (D_\mu - \overset{\leftarrow} {D_\mu}) H$  and $H^\dagger i  \overset{\leftrightarrow}{D_\mu^i} H := i H ^\dagger (\tau^i D_\mu - \overset{\leftarrow} {D_\mu}\tau^i) H$.
\label{tab:dim_six_op}}
\end{center}
\end{table}

\section{Electroweak baryogenesis with higher dimensional operators}
\label{sec:ewbg_eft}
\subsection{Constructing new CPV sources with higher dimensional operators}

When the Higgs field develops a space-time varying vacuum expectation value, $v(x)$, there are operators which interfere with the standard top quark vev insertion diagram to give exotic new sources of $CP$ violation.  We use the closed time path (CTP) formalism \cite{Martin:1959jp,Schwinger:1960qe, Keldysh:1964ud, Chou:1984es, Mahanthappa:1962ex} to calculate $CP$ violating source terms for two operators which facilitate resonantly enhanced $CP$ violating interactions with the bubble wall.  

We also use the vev-insertion approximation (VIA) where the BAU production is dominated by physics in front of the advancing bubble wall. THis is valid when the vev is small compared to both the nucleation temperature and the mass splitting of particles that produce the resonant CPV sources.
%
%
%
%
%
The effective degrees of freedom are then those belonging in the mass eigenbasis of the symmetric (unbroken) phase. Their interactions with the space-time varying vevs are  treated perturbatively under such approximation.  One could perform a resummation to all orders in the vevs following the techniques in \cite{Cirigliano:2009yt, Cirigliano:2011di}.  As a simplification, we ignore the hole modes in the quark plasma \cite{Weldon:1989ys,Weldon:1999th,Klimov:1981ka}.  The effects of mixing with multiparticle states in the thermal bath as well as resummation will also be left to a later, more precise numerical study.  Under these assumptions, the quark propagator reads
\begin{equation}
  S^\lambda (x-y) = \int \frac{d^4 k}{(2 \pi )^4} e^{-i k \cdot (x-y)} g_F ^\lambda (k_0 , \mu _ {t_{L/R}}) \rho ( k_0 ,k) (\cancel{k}+m) ,
\end{equation}
where $\rho(k_0,k)$ is the density of states and 
\begin{equation}
\begin{aligned}
    g_F^>(x)&=1-n_F(x) , \\
    g_F^<(x)&=-n_F(x) ,
\end{aligned}
\end{equation}
with $n_F(x)=(e^{\beta x}+1)^{-1}$.

We will consider two exotic operators, $\mathcal{O}_{t11}$ and $\mathcal{O}_{DD}$. The first can be treated in the usual way by defining the self energy as  
\begin{equation}
  \Sigma_{\text{tot}}  (x,y) = \left( y_t v(x) +\frac{c_i}{\Lambda ^2} v(x)^3 \right) \left(y_t^* v(y) + \frac{c_i ^*}{\Lambda ^2} v(y) ^3\right) S_{t_R}(x,y)\ ,
\end{equation}
whereas the $\mathcal{O}_{DD}$ term has a derivative coupling. For simplicity we will ignore interactions with gauge bosons. Making the replacements $H(x)\rightarrow v(x)$, the self energy is
\begin{equation}
 \Sigma _{\text{tot}} (x,y) = \left(y_{t} v(x) + \frac{c_i}{\Lambda ^2 }\partial _\mu \partial ^\mu v(x)\right   ) \left( y_{t} ^* v(y) +\frac{c_i^*}{\Lambda ^2 }\partial _\mu \partial ^\mu  v(y) \right)S_{t_R}(x,y) .
\end{equation}

The $CP$ conserving term to lowest order in $\Lambda ^{-1}$ for both operators is just the usual resonant relaxation term arising from interactions between the top and the space-time varying vacuum.  The term $v(x)v(y)$ is then expanded near $y=x$ taking the lowest order term. In this case the lowest order is the zeroth order and we find
\begin{equation}
  \begin{aligned}
\Gamma _t &= N_C \frac{|y _t|^2}{2 \pi ^2 T} v(x)^2 \int _0 ^\infty \frac{k^2 dk}{\omega _L \omega _R} {\rm Im} \Bigg[
  \left( \mathcal{E}_L \mathcal{E}_R + k^2 \right)   \left( \frac{h_F(\mathcal{E}_L)+ h_F(\mathcal{E}_R)}{\mathcal{E}_L+\mathcal{E}_R} \right)\\
&  -\left( \mathcal{E}_L \mathcal{E}_R^* - k^2 \right) \left( \frac{h_F(\mathcal{E}_L)+ h_F(\mathcal{E}_R^*)}{\mathcal{E}_R^*-\mathcal{E}_L} \right)\Bigg] .
  \end{aligned}
\end{equation}

\subsection{Contributions from $\mathcal{O}_{t1}$ vertices}
The $CP$ conserving relaxation term up to $O(\Lambda ^{-2})$ just produces the following correction to the Standard Model 
\begin{equation}
  \Gamma _t \mapsto \left(1+ \left| \frac{c_i}{\Lambda ^2} \right|v(x)^2 \right)\Gamma _t \ .
\end{equation}
For the new $CP$ violating source we expand to first order in $z=x$.  The result is
\begin{equation}
{\rm Im } \left[ \frac{c_i y_t ^*}{ \Lambda ^2}  \right] \left[ v(x)^3 v(y) - v(x) v(y) ^3 \right] \mapsto {\rm Im } \left[ \frac{c_i y_t ^*}{ \Lambda ^2}  \right] (z-x)^\mu v(x) ^3 \partial _\mu v(x) \ .
\end{equation}
Only the zeroth component contributes under the assumption of spatial isotropy. Let us also ignore the bubble wall curvature and work in the rest frame of the bubble wall $z= |v_w t-x|$. The time derivative of the vev profile is then a spatial derivative times the wall velocity. In line with the VIA, we assume that the variation of bubble wall with respect to $z$ is sufficiently gentle near the phase boundary \cite{Lee:2004we}. Solving the contour integrals we find
\begin{equation}\label{eq:def_I}
  \begin{aligned}
    S_{\mathcal{O}_{DD}}^{\cancel{CP}} & =  2 \frac{v_w N_C}{\pi ^2} {\rm Im} \left[ \frac{c_i y_t ^*}{\Lambda ^2} \right] v(x) ^3 v^\prime (x)
   \int _0 ^\infty \frac{k^2 dk}{ \omega _L \omega _R} {\rm Im} \Bigg[
   \left( \mathcal{E}_L \mathcal{E}_R + k^2 \right) \left( \frac{n_f(\mathcal{E}_L) - n_F(-\mathcal{E}_R)}{(\mathcal{E}_L+\mathcal{E}_R)^2} \right)\\
  &+ \left( \mathcal{E}_L \mathcal{E}_R ^* - k^2 \right) \left( \frac{n_f(\mathcal{E}_L) - n_F(\mathcal{E}_R ^ *)}{(\mathcal{E}_R^*-\mathcal{E}_L)^2} \right) \Bigg]  \\ &=  2 \frac{v_w N_C}{\pi ^2} {\rm Im} \left[ \frac{c_i y_t ^*}{\Lambda ^2} \right] v(x) ^3 v^\prime (x) I\left[m_{t_L},m_{t_R},\Gamma _{t_R}, \Gamma _{t_L} , \Lambda \right]\ ,
  \end{aligned}
\end{equation}
where we have implicitly defined the function $I[\cdot ]$ 
for notational convenience.

\subsection{Contributions from $\mathcal{O}_{DD}$ vertices}

The $\mathcal{O}_{DD}$ operator requires some care since it involves a derivative coupling to the Higgs.  Once again, we replace the Higgs field with a space-time varying vacuum and expanding the vacuum near $z=x$.  The correction to the SM $CP$ conserving relaxation term comes from the zeroth order term in the expansion
\begin{equation}
  \Gamma _t \mapsto \left(1+ \left| \frac{c_i}{\Lambda ^2 v(x)} \right|v^{\prime \prime}(x) \right) \Gamma _t \ .
\end{equation}
Note that the correction to the relaxation term involves the second derivative of the vev. The usual practice in solving these transport equations is to linearise the differential equations which means assuming the relaxation terms are a constant value in the broken phase. There are some ambiguity in this procedure in that the correction to the above relaxation term varies quite rapidly with $x$ when $x \lesssim L_w$ before going to zero. We therefore linearise the transport equations by setting this correction to its average value between $[0,L_w]$. This will be a somewhat a conservative assumption as this correction will not relax the number densities at all far from the bubble wall.

The $CP$ violating source term, involving the third derivative of the Higgs coming from the next to leading order expansion around $z=x$, is given by
\begin{equation}
  \begin{aligned}
    S_{\mathcal{O}_{DD}}^{\cancel{CP}} & =   \frac{v_w N_C} {\pi ^2}  {\rm Im} \left[ \frac{c_i y_t ^*}{\Lambda ^2} \right]  \left[ v^{\prime \prime \prime }(x)v(x)-  v^{\prime \prime }(x) v^\prime (x) \right]I\left[m_{t_L},m_{t_R},\Gamma _{t_R}, \Gamma _{t_L} , \Lambda \right] \ .
  \end{aligned}
\end{equation}
The derivative coupling causes the operator to be much more  sensitive to the bubble width than the $CP$ violating sources arising from $\mathcal{ O}_{t1}$, or two Higgs doublet models which all have the $CP$ violating source controlled by the first derivative of the vev.  We note that there is a danger that the VIA approximation becomes cruder for derivative couplings particularly when the bubble wall becomes very thin.  Nonetheless, we expect the qualitative result that the source has an increased sensitivity to the wall width to be true even if one uses Wigner functional methods, as it comes from the derivative coupling to the space-time varying vev itself, rather than our approximation scheme. 
\subsection{Calculating the baryon asymmetry}

When calculating the BAU, we make the usual assumption that gauge interactions are very fast and that in the VIA the chemical potential for the $W^\pm$ bosons vanishes as in the symmetric phase.  We ignore interactions with particle species whose interactions are suppressed by small coupling constants.  Specifically, the number densities we consider are the following linear combinations
\begin{equation}
  \begin{aligned}
      Q&= n_{t_L}+n_{b_L} , \\
      T&= n_{t_R} , \\
      H&= n_{H^+} +n_{H^0} .
  \end{aligned}
\end{equation}
Systematically calculating the sources for each self energy term involving the above particle species leads to a network of coupled transport equations. Using the usual relationship $n_i=k_i\mu _iT^2/6$ we can then relate the chemical potentials to the number densities. For operator $\mathcal{ O}_X$ with $X \in \{ DD,t1 \}$ these are
\begin{equation}
  \begin{aligned}
      \partial _\mu Q^\mu & =\Gamma _M \left( \frac{T}{k_T}-\frac{Q}{k_Q} \right) + \Gamma _Y\left(\frac{T}{k_T}- \frac{Q}{k_Q}-\frac{H}{k_H} \right) -2 \Gamma _\text{SS} \mathcal{U} _5  -S ^{\cancel{CP}} _{\mathcal{ O}_X} , \\
      \partial _\mu T^\mu &= -\Gamma _M \left( \frac{T}{k_T}-\frac{Q}{k_Q} \right)-\Gamma _Y\left(\frac{T}{k_T}- \frac{Q}{k_Q}-\frac{H}{k_H} \right)+\Gamma _{\rm SS} \mathcal{ U} _5 +S ^{\cancel{CP}} _{\mathcal{ O}_X} , \\
      \partial _\mu H^\mu &= \Gamma _Y\left(\frac{T}{k_T}- \frac{Q}{k_Q}-\frac{H}{k_H}\right) ,
  \end{aligned}
\end{equation}
where 
\begin{equation}
  \mathcal{U} _5=\left( \frac{2 Q}{k_Q}- \frac{T}{k_T}+\frac{9(Q+T)}{k_B} \right) , \ 
\end{equation}
and the three body Yukawa rates, $\Gamma _Y$, are derived in reference \cite{Cirigliano:2006wh}.  Neglecting the bubble wall curvature we can reduce the problem to a one dimensional one by changing variables to the rest frame of the bubble wall $z=|v_wt-x|$.  We then use the diffusion approximation to write $\nabla \cdot \mathbf{J} = \nabla^2 n$ thus reducing the problem to a set of coupled
differential equations in a single space-time variable.  We do not use the usual simplification that the strong sphaleron and three body Yukawa rates are fast compared to a diffusion time as it has been shown that this assumption can cause an underestimate of the baryon asymmetry in an example model (the MSSM) by a factor of $O(100)$.  While such an analysis has not been done in the SM$+X$, we consider it worth solving the transport analytically using the techniques in \cite{White:2015bva}.
 In the broken phase the solution is
\begin{equation}
   X(z) = \sum_{i=1}^{6} x_1 A_X(\alpha _i) e^{-\alpha_i z}
    \left(\int_0^z \text{d}y, e^{-\alpha_i y} S_{\mathcal{O}_X}^{\cancel{CP}} (y)\right) ,
\end{equation}
and in the symmetric phase we have
\begin{equation}
 X(z) = \sum_{i=1}^{6} A_{X,s} y_1 e^{\gamma_i z} , 
\end{equation}
where $ X \in \left\{Q, T, H \right\}$.
The procedure for how to derive $\alpha _i, \beta _i, x_i,y_i$ and $A_X(\alpha
_i)$ is given in~\cite{White:2015bva}. From these solutions one can then define the left handed number density $n_L(z) = Q_{1L} + Q_{2L} + Q_{3L} = 5Q + 4T$.
The baryon number density, $\rho_B$, satisfies the equation~\cite{Carena:2002ss, Cline:2000nw}
\begin{equation}
  D_Q \rho_B''(z) - v_w \rho_B'(z) - \Theta(-z) \mathcal{ R} \rho _B
   = \Theta(-z) \frac{n_F}{2} \Gamma_{ws} n_L(z) ,
\end{equation}
where $n_F$ is the number of fermion families.  The relaxation parameter is given
by
\begin{equation}
  \mathcal{R} = \frac{15}{4} \Gamma_{ws} ,
\end{equation}
where 
$\Gamma_{ws} \approx 120\,\alpha_W^5 T$ \cite{Bodeker:1999gx, Moore:1999fs, Moore:2000mx}. The baryon asymmetry of the universe, $Y_B$ is then given by
\begin{equation}
  Y_B = -\frac{n_F \Gamma_{ws}}{2\kappa_+ D_Q S} \int _{-\infty}^0
  e^{-\kappa_ - x}\, n_L(x) \,\mathrm{d}x \label{BAU} ,
\end{equation}
where 
\begin{equation}
  \kappa_\pm = \frac{v_w \pm \sqrt{v_w^2 + 4 D_Q \mathcal{ R}}}{2 D_Q} ,
\end{equation}
and the entropy density is
\begin{equation}
  s = \frac{2 \pi ^2}{45} g_* T^3 .
\end{equation}

\section{EDM constraints }\label{subsec:edm}\label{sec:edms}




New sources of $CP$-violation in the Higgs sector are necessary to realise electroweak baryogenesis.  These sources, however, are severely constrained via their contributions to the electric dipole moments (EDMs) of electron, neutron, molecules and atoms.  A direct connection between EDMs and electroweak baryogenesis have been suggested in 
\cite{Fuyuto:2015ida, Huber:2006ri}. 
The sensitivity of these low energy observables owes to contributions from operator mixing and threshold corrections as high scale physics is run down and integrated out.  The present experimental constraints are summarised in Tab.~\ref{tab:edm_constraints}, showing that the electron EDM gives the most stringent bound since it is weakly sensitive to hadronic uncertainties.  This bound is obtained from measurements using polar molecule thorium monoxide (ThO) \cite{Baron:2013eja}.  We therefore focus on contributions to electron EDMs (eEDM) and delay a more comprehensive and systematic treatment to a future study that will include other dimension six operators (cf. e.g. \cite{Cirigliano:2016nyn, Cirigliano:2016njn, Bian:2014zka, Brod:2013cka,Crivellin:2013hpa}).  

\begin{figure}[h!]
\vspace{4ex}	\center
\unitlength=2.2mm
\begin{fmffile}{edm1}
	\begin{fmfgraph*}(30,20)
		\fmfstraight
		\fmfpen{1.2}
		\fmfbottom{f1,f2}
		\fmftop{f3,f4}
		\fmf{vanilla}{f1,v1}
		\fmf{fermion,tension=0.5,label=$f$}{v1,v2}
		\fmf{vanilla}{v2,f2}
		\fmf{phantom}{f3,vb}
		\fmf{phantom,tension=0.5}{va,vb}
		\fmf{phantom}{va,f4}
		\fmffreeze
		\fmf{dashes,label=$h$}{v1,v3}
		\fmf{boson,label=$\gamma$}{v4,v2}
		\fmf{phantom}{vb,v5}
		\fmf{boson,label=$\gamma$}{va,v6}
		\fmfpoly{smooth,tension=0.6}{v3,v4,v6,v5}
		\fmffreeze
		 \fmf{phantom,label.dist=0.1mm,label=$t$}{v3,v5}
	\end{fmfgraph*}
\end{fmffile}
\\[4ex]
\caption{Two-loop Barr-Zee diagrams contributing to the electron EDM.} \label{fig:edm}
\end{figure}
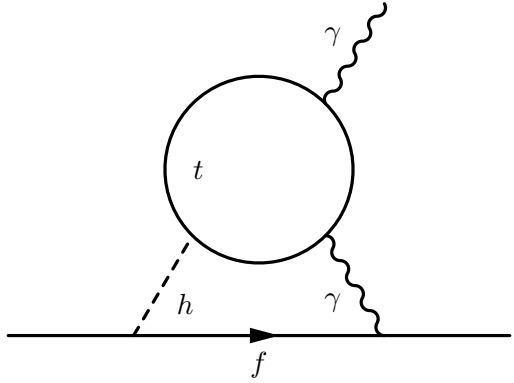

The dipole moment $d_{\psi}$ corresponding to a charged fermion $\psi$ is identified as the coefficient of the five dimensional operator in the effective Lagrangian
\begin{equation}
  \mathcal{L}_{\text{EDM}} = -i d_f \overline{\psi} \gamma^5 \sigma^{\mu\nu} \psi F_{\mu\nu} .
  \label{eq:edm_lagrangian}
\end{equation}
\begin{table}[!h]
  \begin{center}\renewcommand{\arraystretch}{1.3}
    \begin{tabular}[c]{c c c l }
      \hline
       Type & Molecule/Atom &Bounds&\\
       \hline
       \rowcolor{gray!20}
       Paramagnetic & $^{205}$Tl & $|d_\text{Tl}|_{\ }< 1.6\times 10^{-27}\ e$ cm &\cite{Regan:2002ta} \\
       Diamagnetic  & $^{199}$Hg & $|d_\text{Hg}|< 6.2 \times 10^{-30}\ e$ cm &\cite{Griffith:2009zz}\\
       \rowcolor{gray!20}
       Neutron      & $n$        & $|d_{n} |_{\hspace{1ex} }< 3.0\times 10^{-26}\ e$ cm &\cite{Afach:2015sja,Baker:2006ts} \\
       Electron (ThO)     & $e$        & $|d_{e} |_{\hspace{1ex} }< 8.7\times 10^{-29}\ e$ cm & \cite{Baron:2013eja}\\ 
       \hline
     \end{tabular} 
  \end{center}
  \caption{Current limits on electric dipole moments of the electron ($e$), neutron ($n$), mercury ($^{199}$Hg) and thallium ($^{205}$Tl) atoms at 90\% C.L.}     \label{tab:edm_constraints} 
\end{table}
As argued before, one is led to focus on the top-Higgs sector in electroweak baryogenesis due to the $\mathcal{O}(1)$ coupling.  At the non-derivative level, $CP$ violation interactions of this sort are encoded in 
\begin{equation}\label{eq:cpv_tophiggs}
  \begin{aligned}
    \mathcal{L}&\supset -m_t \overline{t}_L t_R - \frac{y_t}{\sqrt{2}} e^{i\xi}h  \overline{t}_L t_R+ h.c.,\\
    &= -m_t \overline{t}_L t_R -  \frac{y_t}{\sqrt{2}} \overline{t} t h \left( \cos\xi + i \gamma^5\sin\xi   \right),
    \end{aligned}
\end{equation}
where $t_L,t_R,h$ are assumed to be in their mass eigenstate and  $m_t=173$ GeV is the physical top mass.  In addition, $y_t$ parametrises the magnitude of the top-Higgs coupling, and $\xi$ its $CP$ phase.  In the SM one has $y_t=y_t^{SM}:=\sqrt{2} m_t/v$ and $\xi=0$.  If there is $CP$ violation in the top-Higgs coupling ($\xi\neq0$) it induces contributions to $d_e$ via two-loop Barr-Zee type diagram \cite{Barr:1990vd} as shown Fig.~\ref{fig:edm}. 
Such contribution is given by\footnote{See \cite{Yamanaka:2014mda} (based on \cite{Yamanaka:2012hm}) for a more pedagogical discussion of the derivation.} 
\begin{equation}\label{eq:edm}
  \frac{d_e}{e}=\frac{16}{3}\frac{\alpha}{(4\pi)^3}\frac{m_e}{y_t^{SM} y_e^{SM}v^2}\left[ y_e^S y_t^P f_1\left( \frac{m_t^2}{m_h^2} \right) + y_e^P y_t^S f_2\left( \frac{m_t^2}{m_h^2} \right) \right] ,
\end{equation}
where the loop functions $f_{1,2}$ are defined in \cite{Stockinger:2006zn, Brod:2013cka}. 
%
We add that other degrees of freedom (not present in our analysis), e.g. charged Higgs boson, may interact with the top quark to give sizable contribution to the EDM via the same the Barr-Zee type diagram. This have been studied in detail in the context of two Higgs doublet models \cite{Leigh:1990kf,Abe:2013qla,Jung:2013hka,BowserChao:1997bb}.

Firstly we discuss how the $\mathcal{O}_{t1}$ operator leads to $CP$ violating top-Higgs coupling of the form (\ref{eq:cpv_tophiggs}) by expansion of the $H$ operator around its vev.  With $H=\frac{1}{\sqrt{2}}( 0, v+h)^T$, this leads to
\begin{equation}
  \begin{aligned}
    \mathcal{L}& \supset -\left( \alpha + \frac{c_{t1}}{\Lambda^2} H^\dagger H  \right)\overline{Q}_L\tilde{H} t_R + h.c. \\
    &=-\underbrace{\frac{1}{\sqrt{2}} \left(\alpha+c_{t1}\frac{v^2}{\Lambda^2}\right) v}_{m_te^{i\xi_m}} \overline{t}_L t_R
    -\underbrace{\left( \alpha +3c_{t1}\frac{v^2}{\Lambda^2} \right)}_{y_te^{i\xi_t}}\frac{h}{\sqrt{2}}\overline{t}_L t_R + h.c. .
  \end{aligned}
\end{equation}
These operators are brought into their mass basis by a field redefinition $t_R \mapsto e^{-i\xi_m} t_R $.  In such case, the physical $CP$ phase can be identified with $\xi_t-\xi_m$. 

In case of the $\mathcal{O}_{DD}$ operator, the top-Higgs interaction contains a derivative.  In principle, this contributes to $d_e$ through the same two-loop diagram, shown in Fig.~\ref{fig:edm}, and one can derive an analogue of (\ref{eq:edm}) with the momentum dependent top-Higgs vertex.  Differing from the discussion of the baryon production during the EWPT, the Higgs vev here corresponds to the one well after the EWPT and is hence not space-time dependent. It is valid then to use classical EOMs to recast $\mathcal{O}_{DD}$ in terms of derivative free operators as in equation (\ref{eq:ODD_EOM}). 

The dominant constraints on $\mathcal{O}_{DD}$ come from from the first term of (\ref{eq:ODD_EOM}), since four-fermion operators to do not lead to sensitive observables  \cite{Cirigliano:2016nyn}.  Following the previous steps, one obtains 
\begin{equation}
  \begin{aligned}
    \mathcal{L}& \supset -\left[ \alpha +  \frac{c_{DD} }{\Lambda^2} (\mu^2 -\lambda H^\dagger H)\right]\overline{Q}_L\tilde{H} t_R + h.c. \\
    &=-\underbrace{ \frac{1}{\sqrt{2}}\left(\alpha +\frac{c_{DD}}{\Lambda^2}
      \left( \mu^2 - \frac{1}{2}\lambda v^2 \right)
    \right)v }_{m_te^{i\xi_m}} \overline{t}_L t_R
    -\underbrace{\left[ \alpha +\frac{c_{DD}}{\Lambda^2}\left( \mu^2 - \frac{3}{2}\lambda v^2 \right)\right]}_{y_te^{i\xi_t}}\frac{h}{\sqrt{2}}\overline{t}_L t_R + h.c. .
  \end{aligned}
\end{equation}

\begin{figure}[h!]
  \begin{subfigure}[h]{0.49\textwidth}\center
    \includegraphics[trim=4cm 3mm 1cm 20mm ,clip,width=\textwidth]{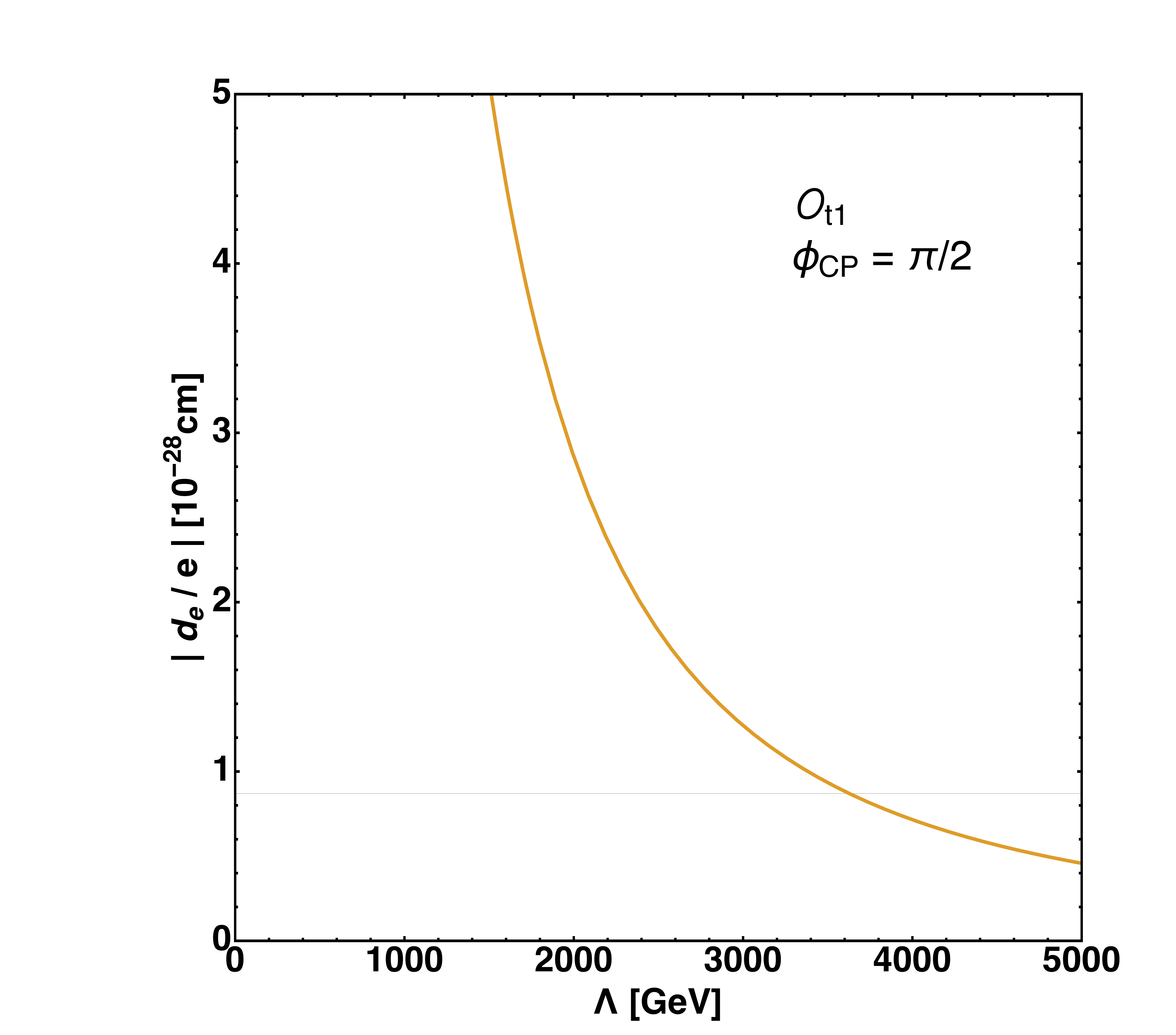}
    \caption{}
  \end{subfigure}
  \begin{subfigure}[h]{0.49\textwidth}\center
    \includegraphics[trim=4cm 3mm 1cm 20mm ,clip,width=\textwidth]{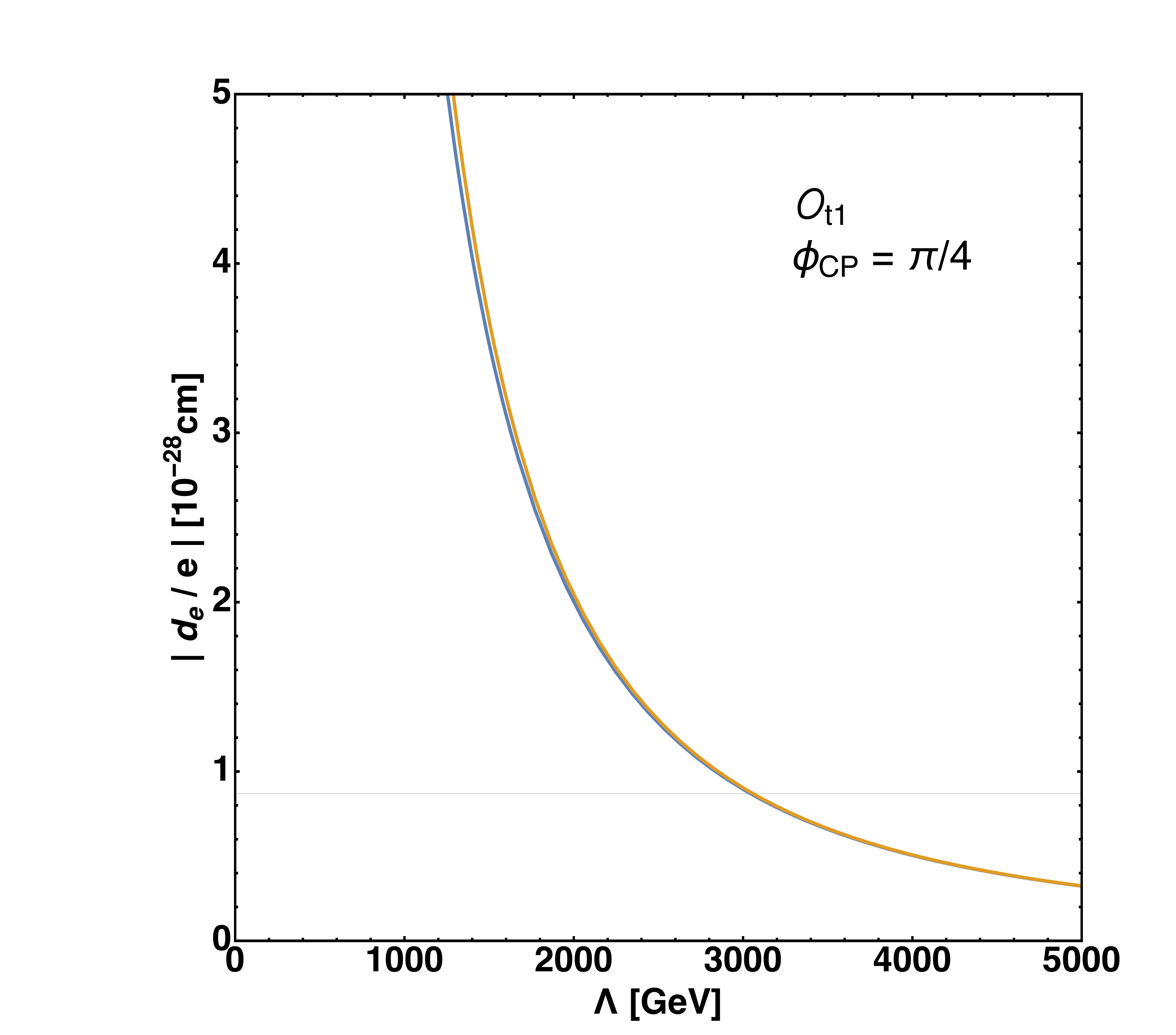}
    \caption{}
  \end{subfigure}

  \begin{subfigure}[h]{0.49\textwidth}\center
    \includegraphics[trim=4cm 3mm 1cm 20mm ,clip,width=\textwidth]{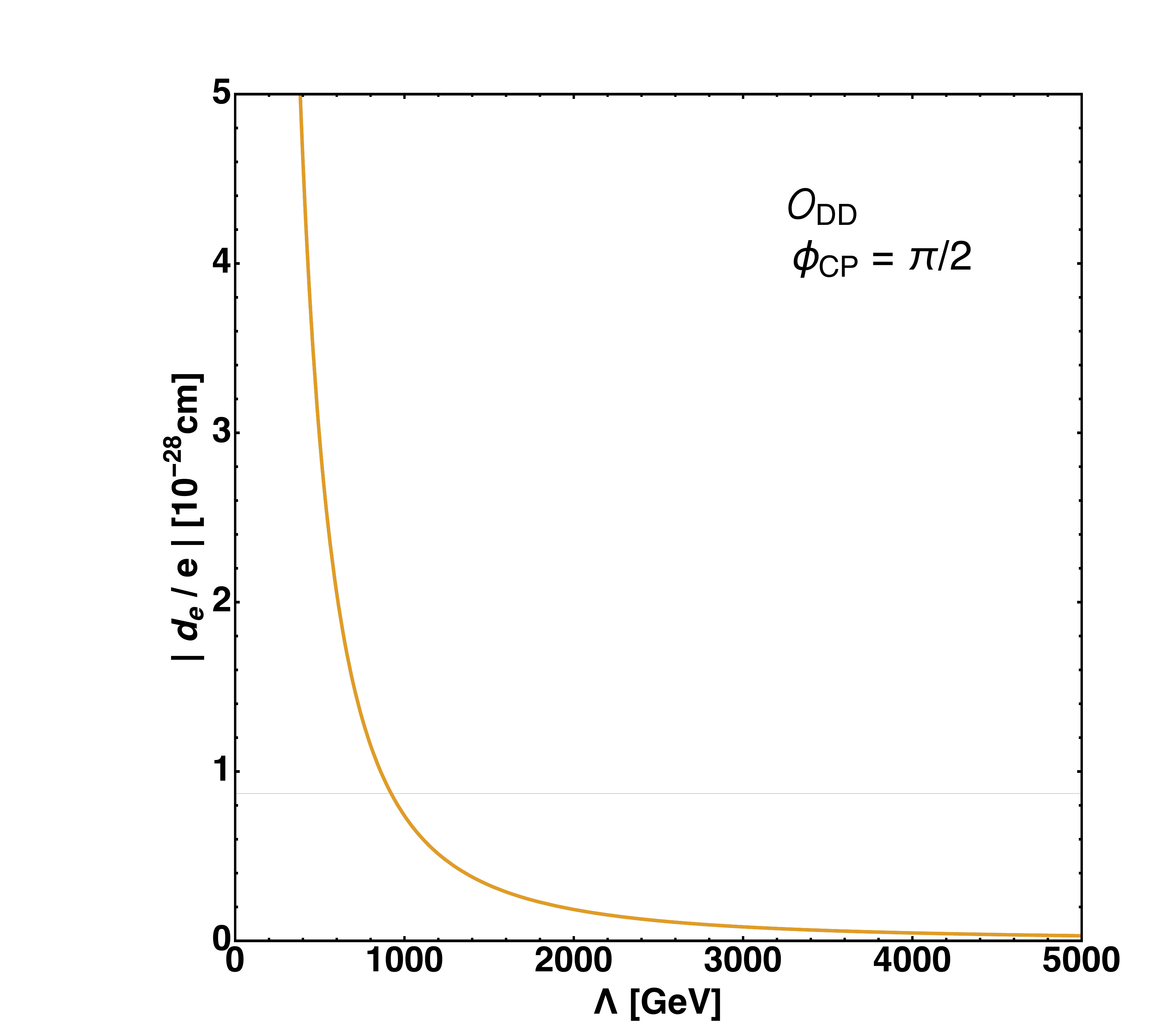}
    \caption{}
  \end{subfigure}
  \begin{subfigure}[h]{0.49\textwidth}\center
    \includegraphics[trim=4cm 3mm 1cm 20mm ,clip,width=\textwidth]{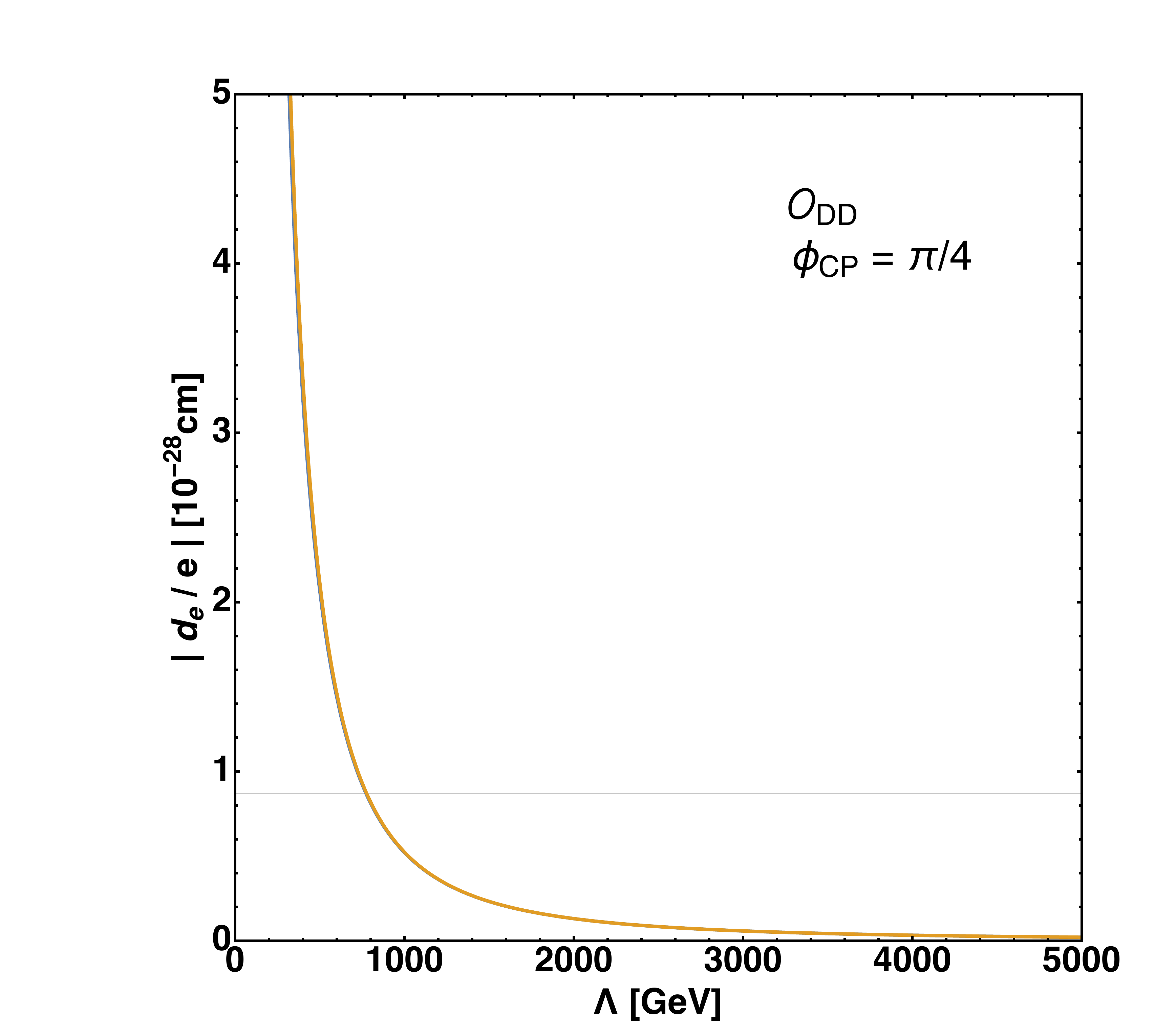}
    \caption{}
  \end{subfigure}
  \caption{Two loop contribution to the electron electric dipole moment via a top quark due to the $\mathcal{O}_{t1}$ and $\mathcal{O}_{DD}$ operators. Here $\phi_{CP}$ denotes the phase $c_{DD,t1}=e^{i\phi_{CP}}$ of the Wilson coefficient appearing in front of the operator.  The horizontal line corresponds to the experimental limit. 
  }
  \label{fig:edm_bounds}
\end{figure}

In both of these cases, one assumes a generic coefficient $\alpha\in \mathbb{C}$ for the dimension-four top Yukawa coupling $ \overline{Q}_L \tilde{H} t_R$. 
Making the assumption that $CP$-violation comes only from the $d=6$ operators and that the scale of the operator is set by the cutoff, one sets  $\text{Im} \left(\alpha\right) =0$ and $c_{DD,t1} = e^{i\phi_{CP}}$. 
The value of $\alpha $ is chosen to absorb the effects of the $\mathcal{O}_{DD,t1}$ and to reproduce $m_t = 173$ GeV.  
Currently, we take $\mu^2=m_h^2$ 
and $\mu^2 =\lambda v^2$ but we note that this relation can be modified by pure Higgs effective operators such as $\left(H^\dagger H\right)^3$.  Fig.~\ref{fig:edm_bounds} shows the contributions of the $\mathcal{O}_{t1}$ and $\mathcal{O}_{DD}$ operators to the electron EDM as a function of the cutoff scale $\Lambda$. For the former, operator a strong dependence on the $CP$ phase of the higher dimensional operator is observed.  Particularly, a cutoff of $\Lambda \gtrsim 3600$ GeV is required to remain consistent with the current constraints $\phi_{CP}=\pi/2$, but is relaxed to  $\Lambda \gtrsim 3000$ GeV for $\phi_{CP}=\pi/4$. The electron EDM bound on the latter operator is weaker, with the cutoff scale roughly required to be $\Lambda \gtrsim 1$ TeV for both $CP$ phases.  One should keep in mind that when interpreting these results, one assumes a pure scalar electron Yukawa coupling with its SM value (cf. \cite{Altmannshofer:2015qra} and references therein for discussions on experimental constraints of such coupling). 

\section{Numerical results and discussion}\label{sec:results}

\begin{figure}[h!]
    \includegraphics[trim=15mm 5mm 2mm 10mm ,clip,width=\textwidth]{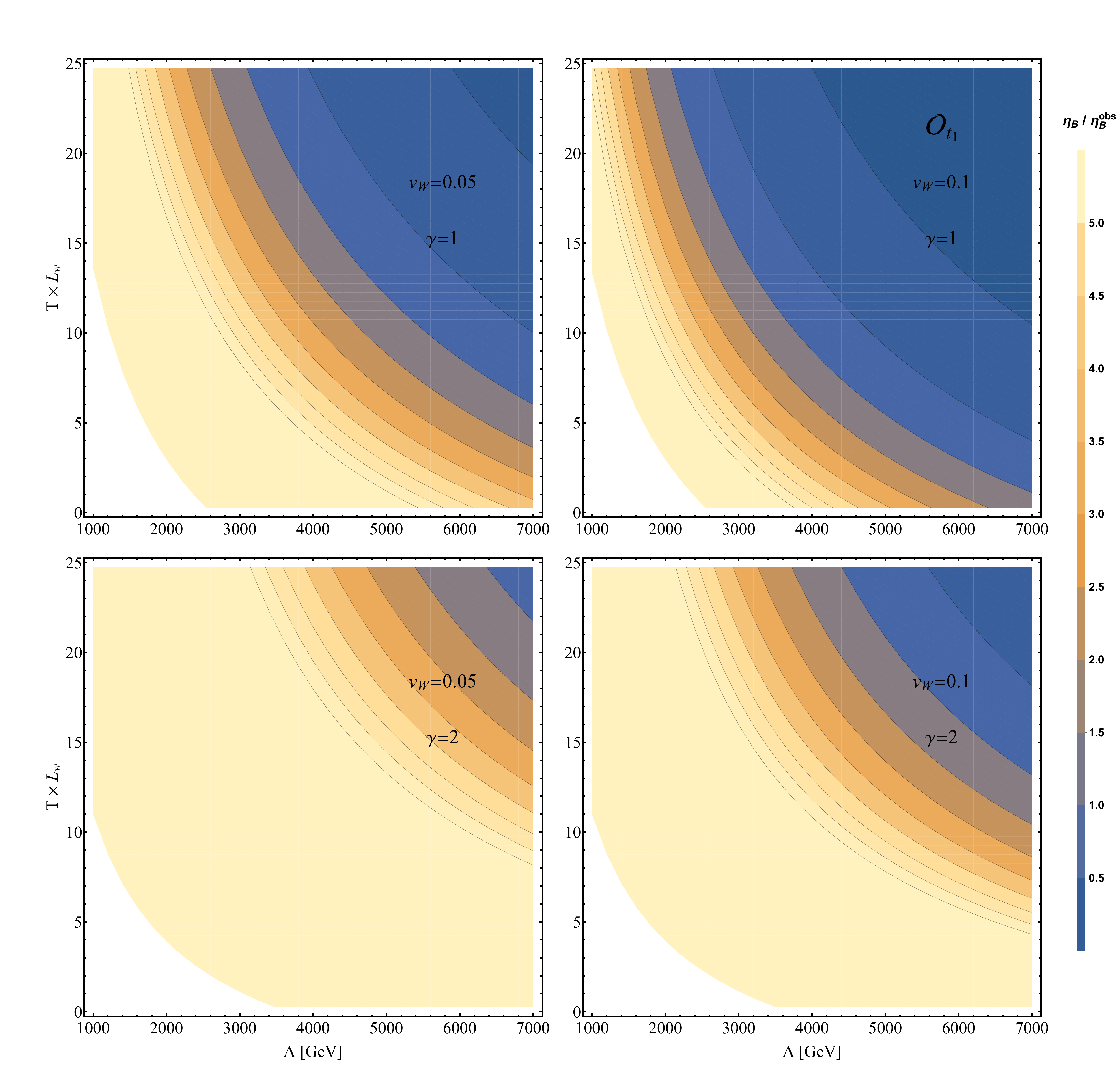}
    \caption{The baryon asymmetry due to the $CP$ violating operator $\mathcal{O}_{t1}$ in the plane of the bubble wall width vs. cutoff $(L_w, \Lambda)$. The dependence on $L_w$ is relatively gentle.}\label{fig:Ot1}
\end{figure}
\begin{figure}[h!]
    \includegraphics[trim=15mm 5mm 2mm 10mm ,clip,width=\textwidth]{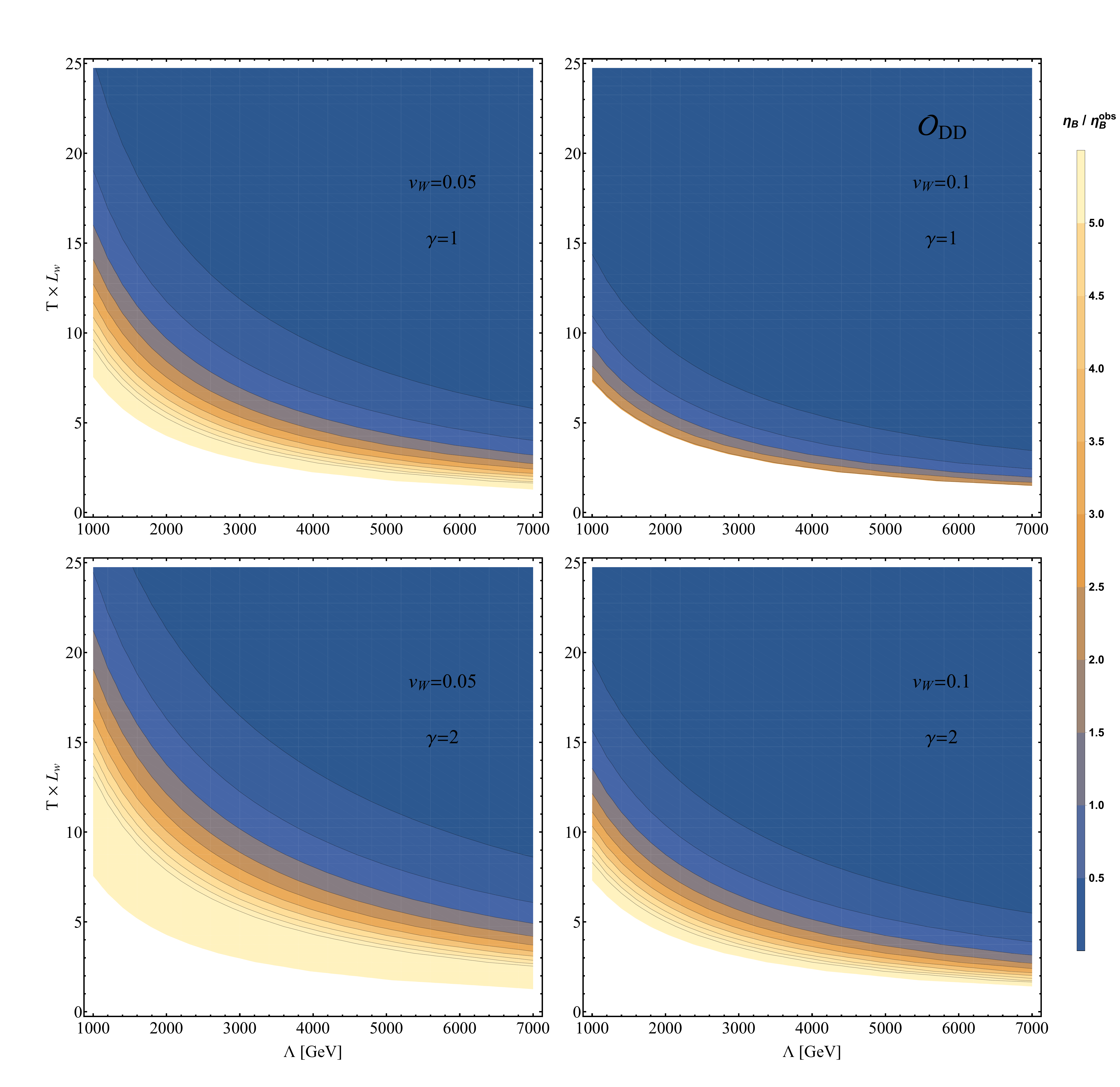}
    \caption{The baryon asymmetry due to the $CP$ violating operator $\mathcal{ O}_{DD}$ in the plane of the bubble wall width vs. cutoff $(L_w, \Lambda)$. The dependence on $L_w$ is quite steep. }\label{fig:ODD}
\end{figure}

We plot the BAU produced by the new $CP$ violating sources resulting from the operators $\mathcal{ O}_{DD}$ and $\mathcal{ O}_{t1}$ in Fig.~\ref{fig:Ot1}  and Fig.~\ref{fig:ODD} respectively. We set the nucleation temperature to $T_n=100$ GeV and the $CP$ violating phase $\phi_{CP}=\pi/2$ such that new coupling constants are $c_{DD,t1}=i$ (cf. Sec.~\ref{sec:edms}). We then set the value of the vev deep within the broken phase to obtain two different values of the order parameter $\gamma := v(T)/T$. The first value is the minimal value of unity --- since this is the approximate condition for a strongly first order phases transition necessary to sufficiently suppress sphaleron interactions deep in the broken phase thereby preserving the baryon number. The higher value is $\gamma =2$ since this is an approximate maximum value for $\gamma$ during the electroweak phase transition for a critical temperature $T_c\geq 100$ \cite{Profumo:2007wc}.  Generically, a smaller value of the wall velocity produces a larger BAU as does a larger value of $\gamma$ which is expected given that the $CP$ violating sources are all proportional to $\gamma$ to some power. One should note that the Standard Model with a light Higgs has a larger wall velocity. The wall velocity can be suppressed by additional particles in the plasma which might also be heavy enough to justify an effective field theory approach. Therefore, we can once again parametrise our ignorance of such particles just by keeping the wall velocity as a free parameter and setting it to values $0.05$ and $0.1$. As explained in Sec.~\ref{sec:edms}, the minimum cutoff for the operator $\mathcal{ O}_{DD}$ is about a TeV whereas the minimum cutoff for the $\mathcal{ O}_{t1}$ operator is significantly higher, about $3.5$ TeV, due to its effect on the top quark Yukawa. 

As expected, the baryon asymmetry due to the operator $\mathcal{ O}_{DD}$ is very sensitive to the bubble wall width. In both cases, a thin bubble wall is favoured with a large proportion of the parameter space already ruled out. However, the baryon asymmetry diverges quickly for very small values of $L_w$ for the operator ${\cal O}_{DD}$. It would be very interesting to see how strongly this effect persists when one goes beyond the VIA by using techniques described in \cite{Cirigliano:2009yt,Cirigliano:2011di}. 

Remarkably the BAU can be produced by the ${\cal O}_{t1}$ operator with extremely large values of the cutoff if the wall width and velocity are small but the order parameter $\gamma$ is large. This is due to the fact that the $CP$ violating source scales as $v( T)^4/\Lambda ^2$ so the suppression due to the cutoff is not as severe as it is for the ${\cal O}_{DD}$ operator. This also means that the BAU for operator ${\cal O}_{t1}$ is more sensitive to the value of $v(T)$. However, explaining the baryon asymmetry with the $\mathcal{ O}_{t1}$ operator is not viable with about a 6-fold increase in the minimal value of $\Lambda$ (or equivalently $\Lambda/|c_i|$). This means that this operator may be completely ruled out as a sole explanation to the BAU in the foreseeable future if EDM searches improve in sensitivity by about an order of magnitude or measurements of the top quark Yukawa coupling become moderately more accurate. There is of course the caveat that the baryon asymmetry has some moderate dependence on the nucleation temperature. 

Not all baryon number produced during the electroweak phase transition is preserved until the phase transition is finished. The fraction that is preserved has a double exponential dependence on the strength of the order parameter $v(T_c)/T_c$. So for an order parameter of $v(T_c)/T_c \approx 0.75$ one might need to produce as much as $10$ times of the observed baryon asymmetry \cite{Patel:2011th}. Including the effects of washout, with detailed calculations of the sphaleron energy, we leave to an interesting future project.

\subsection{Space-time dependent cutoff}\label{sec:spacetime_dep}

Within the approach of effective field theory, we approximate the propagators of heavy particles by the inverse of their mass squared. If a particle acquires some of its mass via symmetry breaking, the mass of the heavy particle inherits a space-time dependence via the vev of the other field such that
\begin{equation}
    \frac{1}{\Lambda ^2} \to \frac{1}{\Lambda^2 _0(T)+\Delta \Lambda ^2(x)} := \frac{\kappa(x)}{\Lambda^2 _0(T)+v^2(x)}  \ .
\end{equation}
Here $\kappa (x)$ is a space-time dependent function absorbing the effects of heavy physics which the EFT is ignorant of. An example for such situation is an EFT for sparticles in supersymmetry that have a soft mass but acquire some contribution to their masses from the vacuum expectation value of the Higgs.

In this section, we argue that a space-time dependent cutoff is not necessarily fatal for an effective field theory, although we do not claim that our treatment is comprehensive.  For example, we do not discuss any subtleties that may arise from the fact that the space-time varying cutoff is defined within a particular frame of reference (although comfort ourselves with the fact that temperature is also defined within a particular reference frame).
For the effective field theory to remain valid $\Lambda _0$ has to be high enough to justify the new physics it represents to be heavy enough.  Since CPV source terms generically will depend on the derivative of $\Lambda $, one could ask if it is in principle possible that such a term can boost the baryon asymmetry.  The answer is typically no in the case where $\Lambda _0(T)$ is large enough as corrections to the $CP$ violating source will be of the order 
\begin{equation}
  \frac{1}{\Lambda ^4(x)} \frac{\partial  }{\partial x} \left[ \Delta \Lambda ^2 (x)\right]\ . 
\end{equation} 

Finally, we briefly consider the case where we definitely do not expect effective field theory to work when $\Lambda _ 0(T) \to 0$. Using intuition about the generic behaviour of space-time varying functions during the electroweak phase transition (such as CPV phases, vevs and variation of the ratio of vevs $\beta (x)$) we can make an \textit{ansatz}  to parametrise our ignorance of the new physics
\begin{equation}
  \kappa(x) = \kappa _0 + \frac{\Delta \kappa }{2} \left[1+\tanh\left(\frac{x}{L_w}\right) \right] \ .
\end{equation}
Suppose we have the case where our $D=6$ operator that we test is $\mathcal{O}_{t1}$ which is acquired by integrating out a heavy Higgs in a two Higgs doublet model. If we set $\Lambda _0 $ to zero\footnote{This is physically unrealistic as there is always a thermal mass mass but done for illustrative purposes.} the $CP$ violating source we get is
\begin{equation}
    S_{\mathcal{O}_{t1}} ^{\cancel{CP}}= v^2(x) \dot{\kappa} (x) {\rm Im} [y_t c_i^*] I\left( m_{t_L},m_{t_R},\Gamma _{t_L},\Gamma _{t_R}, \Lambda (x_i) \right) \ .
\end{equation}
Here $\Lambda (x_i)$ is the cutoff evaluated at a single space-time point for simplicity. This is, of course, the most ambiguous part of this discussion.
We can compare the above to the two Higgs doublet model where one gets
\begin{equation}
    S_{2HDM} ^{\cancel{CP}} = v^2(x) \dot{\beta} (x) {\rm Im} [y_{t_1} y_{t_2}^*] I\left( m_{t_L},m_{t_R},\Gamma _{t_L},\Gamma _{t_R} \right) \ .
\end{equation}
Remarkably, the effective field theory framework reproduces much of the structure of the UV complete theory in a case where we had no right to expect this. 

One should not, however, take the comparisons between the above two CPV sources too literally.  If we replaced the cutoff with the mass of the second Higgs doublet we would acquire coefficients with complicated dependence on parameters beyond the SM.  This is expected since the heavy physics that produces $\mathcal{O}_{t1}$ is not unique and the EFT framework is necessarily somewhat ignorant of the UV completion. What is remarkable here is that the EFT framework produces the correct dependence on the masses and thermal widths of the top quark, including resonance effects, the correct dependence on the vev profiles, the top Yukawa coupling as well as the variation of the space-time dependence of the heavy physics all in a scenario where the EFT framework is expected to be crude.  While this may be coincidental, it would be interesting for future work to ascertain how well the effective field theory works in calculating the BAU for a variety of models where $\Lambda _0(T)$ is small.

\section{Conclusions}\label{sec:concl}

The growing sensitivity of electric dipole moment searches is increasingly constraining the parameter space of baryogenesis models.  Consequently, in the near future various electroweak baryogenesis models will be either confirmed or ruled out by EDM searches.  The number of baryogenesis models, however, is rendering the application of experimental bounds (including EDM limits) on each model impractical.  This necessitates a model independent, direct connection between EDM constraints and BAU calculations.  In this work, we studied such a connection using the framework of an effective field theory.  

Examining the connection between dimension six effective operators and the BAU, we found that the conventional degeneracy is broken between operators containing derivatives of the Higgs field and their counterparts related by the equation of motion.  According to the na\"ive CPV analysis, higher order contributions which  arises when derivative operators are traded to non-derivative ones, can be safely neglected since they are suppressed by the cutoff scale.  When calculating the BAU, however, operators containing a derivative of the Higgs filed yield a CPV contribution to baryogenesis that involves the derivative of the Higgs vev.  If one trades these operators to non-derivative ones then one completely changes the nature of the CPV contribution to baryogenesis.  The removal of $\mathcal{O}^{(m)}_{\Delta V,D=4+n}$ due to power-counting arguments in the EOMs when relating $\mathcal{O}_{D=6,\text{CPV}}$ with SM operators is problematic as a rapidly varying Higgs wall profile destabilises the hierarchy between the  vev and cutoff scale.  


After re-classifying dimension six effective operators, we selected two simple dimension six operators (one containing a derivative and the other not) and calculated the respective baryon asymmetry.  We also subjected these operators to EDM constrains, thereby directly connecting the effect of the EDM constraints to the amount of baryon asymmetry these operators can yield.
Finally, we discussed the possibility of the effective cutoff being space-time dependent and showed that the effective field theory approach captures the bulk of the correct physics even when we expect it to be a crude approximation.

We stress that the baryon asymmetry calculated from the normally neglected dimension six operators involving derivative coupling to the Higgs is more sensitive to the bubble dynamics of the EWPT.  The approach we suggest does not apply to more complicated scenarios such as multistep phase transitions (cf. e.g. \cite{Patel:2012pi}).  Also, more work needs to be done analysing these operators using the full Wigner functional approach presented in references \cite{Tulin:2011wi}.  Nonetheless, we have made a step toward a more general test of the electroweak baryogenesis paradigm.


\acknowledgments

We like to thank W. Denkens, J. de Vries, Nodoka Yamanaka, Michael Schmidt and Chuan-Ren Chen for useful discussions. This work was partially supported by the Australian Research Council. JY will like to particularly thank Archil Kobakhidze for supporting this work during his transitional period.


\bibliographystyle{mybibsty}
\bibliography{myrefs}

\providecommand{\href}[2]{#2}\begingroup\raggedright\begin{thebibliography}{10}

\bibitem{Aad:2012tfa}
{\scshape ATLAS} collaboration, G.~Aad et~al., \emph{{Observation of a new
  particle in the search for the Standard Model Higgs boson with the ATLAS
  detector at the LHC}},
  \href{http://dx.doi.org/10.1016/j.physletb.2012.08.020}{\emph{Phys. Lett.}
  {\bf B716} (2012) 1--29}, [\href{http://arxiv.org/abs/1207.7214}{{\tt
  1207.7214}}].

\bibitem{Chatrchyan:2012xdj}
{\scshape CMS} collaboration, S.~Chatrchyan et~al., \emph{{Observation of a new
  boson at a mass of 125 GeV with the CMS experiment at the LHC}},
  \href{http://dx.doi.org/10.1016/j.physletb.2012.08.021}{\emph{Phys. Lett.}
  {\bf B716} (2012) 30--61}, [\href{http://arxiv.org/abs/1207.7235}{{\tt
  1207.7235}}].

\bibitem{White:2016nbo}
G.~A. White, \emph{{A Pedagogical Introduction to Electroweak Baryogenesis}}.
\newblock IOP Concise Physics. Morgan \& Claypool, 2016,
  \href{http://dx.doi.org/10.1088/978-1-6817-4457-5}{10.1088/978-1-6817-4457-5}.

\bibitem{Aad:2015zhl}
{\scshape ATLAS, CMS} collaboration, G.~Aad et~al., \emph{{Combined Measurement
  of the Higgs Boson Mass in $pp$ Collisions at $\sqrt{s}=7$ and 8 TeV with the
  ATLAS and CMS Experiments}},
  \href{http://dx.doi.org/10.1103/PhysRevLett.114.191803}{\emph{Phys. Rev.
  Lett.} {\bf 114} (2015) 191803}, [\href{http://arxiv.org/abs/1503.07589}{{\tt
  1503.07589}}].

\bibitem{Rummukainen:1998as}
K.~Rummukainen, M.~Tsypin, K.~Kajantie, M.~Laine and M.~E. Shaposhnikov,
  \emph{{The Universality class of the electroweak theory}},
  \href{http://dx.doi.org/10.1016/S0550-3213(98)00494-5}{\emph{Nucl. Phys.}
  {\bf B532} (1998) 283--314},
  [\href{http://arxiv.org/abs/hep-lat/9805013}{{\tt hep-lat/9805013}}].

\bibitem{Gavela:1993ts}
M.~B. Gavela, P.~Hernandez, J.~Orloff and O.~Pene, \emph{{Standard model CP
  violation and baryon asymmetry}},
  \href{http://dx.doi.org/10.1142/S0217732394000629}{\emph{Mod. Phys. Lett.}
  {\bf A9} (1994) 795--810}, [\href{http://arxiv.org/abs/hep-ph/9312215}{{\tt
  hep-ph/9312215}}].

\bibitem{Konstandin:2003dx}
T.~Konstandin, T.~Prokopec and M.~G. Schmidt, \emph{{Axial currents from CKM
  matrix CP violation and electroweak baryogenesis}},
  \href{http://dx.doi.org/10.1016/j.nuclphysb.2003.11.037}{\emph{Nucl. Phys.}
  {\bf B679} (2004) 246--260}, [\href{http://arxiv.org/abs/hep-ph/0309291}{{\tt
  hep-ph/0309291}}].

\bibitem{Ade:2015xua}
{\scshape Planck} collaboration, P.~A.~R. Ade et~al., \emph{{Planck 2015
  results. XIII. Cosmological parameters}},
  \href{http://dx.doi.org/10.1051/0004-6361/201525830}{\emph{Astron.
  Astrophys.} {\bf 594} (2016) A13},
  [\href{http://arxiv.org/abs/1502.01589}{{\tt 1502.01589}}].

\bibitem{Kobakhidze:2015xlz}
A.~Kobakhidze, L.~Wu and J.~Yue, \emph{{Electroweak Baryogenesis with Anomalous
  Higgs Couplings}},
  \href{http://dx.doi.org/10.1007/JHEP04(2016)011}{\emph{JHEP} {\bf 04} (2016)
  011}, [\href{http://arxiv.org/abs/1512.08922}{{\tt 1512.08922}}].

\bibitem{Bodeker:2004ws}
D.~Bodeker, L.~Fromme, S.~J. Huber and M.~Seniuch, \emph{{The Baryon asymmetry
  in the standard model with a low cut-off}},
  \href{http://dx.doi.org/10.1088/1126-6708/2005/02/026}{\emph{JHEP} {\bf 02}
  (2005) 026}, [\href{http://arxiv.org/abs/hep-ph/0412366}{{\tt
  hep-ph/0412366}}].

\bibitem{Fromme:2006wx}
L.~Fromme and S.~J. Huber, \emph{{Top transport in electroweak baryogenesis}},
  \href{http://dx.doi.org/10.1088/1126-6708/2007/03/049}{\emph{JHEP} {\bf 03}
  (2007) 049}, [\href{http://arxiv.org/abs/hep-ph/0604159}{{\tt
  hep-ph/0604159}}].

\bibitem{Huang:2015izx}
F.~P. Huang, P.-H. Gu, P.-F. Yin, Z.-H. Yu and X.~Zhang, \emph{{Testing the
  electroweak phase transition and electroweak baryogenesis at the LHC and a
  circular electron-positron collider}},
  \href{http://dx.doi.org/10.1103/PhysRevD.93.103515}{\emph{Phys. Rev.} {\bf
  D93} (2016) 103515}, [\href{http://arxiv.org/abs/1511.03969}{{\tt
  1511.03969}}].

\bibitem{Engel:2013lsa}
J.~Engel, M.~J. Ramsey-Musolf and U.~van Kolck, \emph{{Electric Dipole Moments
  of Nucleons, Nuclei, and Atoms: The Standard Model and Beyond}},
  \href{http://dx.doi.org/10.1016/j.ppnp.2013.03.003}{\emph{Prog. Part. Nucl.
  Phys.} {\bf 71} (2013) 21--74}, [\href{http://arxiv.org/abs/1303.2371}{{\tt
  1303.2371}}].

\bibitem{Chupp:2014gka}
T.~Chupp and M.~Ramsey-Musolf, \emph{{Electric Dipole Moments: A Global
  Analysis}}, \href{http://dx.doi.org/10.1103/PhysRevC.91.035502}{\emph{Phys.
  Rev.} {\bf C91} (2015) 035502}, [\href{http://arxiv.org/abs/1407.1064}{{\tt
  1407.1064}}].

\bibitem{Chien:2015xha}
Y.~T. Chien, V.~Cirigliano, W.~Dekens, J.~de~Vries and E.~Mereghetti,
  \emph{{Direct and indirect constraints on CP-violating Higgs-quark and
  Higgs-gluon interactions}},
  \href{http://dx.doi.org/10.1007/JHEP02(2016)011}{\emph{JHEP} {\bf 02} (2016)
  011}, [\href{http://arxiv.org/abs/1510.00725}{{\tt 1510.00725}}].

\bibitem{Brod:2013cka}
J.~Brod, U.~Haisch and J.~Zupan, \emph{{Constraints on CP-violating Higgs
  couplings to the third generation}},
  \href{http://dx.doi.org/10.1007/JHEP11(2013)180}{\emph{JHEP} {\bf 11} (2013)
  180}, [\href{http://arxiv.org/abs/1310.1385}{{\tt 1310.1385}}].

\bibitem{Noble:2007kk}
A.~Noble and M.~Perelstein, \emph{{Higgs self-coupling as a probe of
  electroweak phase transition}},
  \href{http://dx.doi.org/10.1103/PhysRevD.78.063518}{\emph{Phys. Rev.} {\bf
  D78} (2008) 063518}, [\href{http://arxiv.org/abs/0711.3018}{{\tt
  0711.3018}}].

\bibitem{Delaunay:2007wb}
C.~Delaunay, C.~Grojean and J.~D. Wells, \emph{{Dynamics of Non-renormalizable
  Electroweak Symmetry Breaking}},
  \href{http://dx.doi.org/10.1088/1126-6708/2008/04/029}{\emph{JHEP} {\bf 04}
  (2008) 029}, [\href{http://arxiv.org/abs/0711.2511}{{\tt 0711.2511}}].

\bibitem{Cirigliano:2016njn}
V.~Cirigliano, W.~Dekens, J.~de~Vries and E.~Mereghetti, \emph{{Is There Room
  for CP Violation in the Top-Higgs Sector?}},
  \href{http://dx.doi.org/10.1103/PhysRevD.94.016002}{\emph{Phys. Rev.} {\bf
  D94} (2016) 016002}, [\href{http://arxiv.org/abs/1603.03049}{{\tt
  1603.03049}}].

\bibitem{Cirigliano:2016nyn}
V.~Cirigliano, W.~Dekens, J.~de~Vries and E.~Mereghetti, \emph{{Constraining
  the top-Higgs sector of the Standard Model Effective Field Theory}},
  \href{http://arxiv.org/abs/1605.04311}{{\tt 1605.04311}}.

\bibitem{Grojean:2004xa}
C.~Grojean, G.~Servant and J.~D. Wells, \emph{{First-order electroweak phase
  transition in the standard model with a low cutoff}},
  \href{http://dx.doi.org/10.1103/PhysRevD.71.036001}{\emph{Phys. Rev.} {\bf
  D71} (2005) 036001}, [\href{http://arxiv.org/abs/hep-ph/0407019}{{\tt
  hep-ph/0407019}}].

\bibitem{Fuyuto:2015ida}
K.~Fuyuto, J.~Hisano and E.~Senaha, \emph{{Toward verification of electroweak
  baryogenesis by electric dipole moments}},
  \href{http://dx.doi.org/10.1016/j.physletb.2016.02.053}{\emph{Phys. Lett.}
  {\bf B755} (2016) 491--497}, [\href{http://arxiv.org/abs/1510.04485}{{\tt
  1510.04485}}].

\bibitem{Huber:2006ri}
S.~J. Huber, M.~Pospelov and A.~Ritz, \emph{{Electric dipole moment constraints
  on minimal electroweak baryogenesis}},
  \href{http://dx.doi.org/10.1103/PhysRevD.75.036006}{\emph{Phys. Rev.} {\bf
  D75} (2007) 036006}, [\href{http://arxiv.org/abs/hep-ph/0610003}{{\tt
  hep-ph/0610003}}].

\bibitem{Shu:2013uua}
J.~Shu and Y.~Zhang, \emph{{Impact of a CP Violating Higgs Sector: From LHC to
  Baryogenesis}},
  \href{http://dx.doi.org/10.1103/PhysRevLett.111.091801}{\emph{Phys. Rev.
  Lett.} {\bf 111} (2013) 091801}, [\href{http://arxiv.org/abs/1304.0773}{{\tt
  1304.0773}}].

\bibitem{Zhang:1993vh}
X.~Zhang and B.~L. Young, \emph{{Effective Lagrangian approach to electroweak
  baryogenesis: Higgs mass limit and electric dipole moments of fermion}},
  \href{http://dx.doi.org/10.1103/PhysRevD.49.563}{\emph{Phys. Rev.} {\bf D49}
  (1994) 563--566}, [\href{http://arxiv.org/abs/hep-ph/9309269}{{\tt
  hep-ph/9309269}}].

\bibitem{Huang:2015bta}
F.~P. Huang and C.~S. Li, \emph{{Electroweak baryogenesis in the framework of
  the effective field theory}},
  \href{http://dx.doi.org/10.1103/PhysRevD.92.075014}{\emph{Phys. Rev.} {\bf
  D92} (2015) 075014}, [\href{http://arxiv.org/abs/1507.08168}{{\tt
  1507.08168}}].

\bibitem{Damgaard:2015con}
P.~H. Damgaard, A.~Haarr, D.~O'Connell and A.~Tranberg, \emph{{Effective Field
  Theory and Electroweak Baryogenesis in the Singlet-Extended Standard Model}},
  \href{http://dx.doi.org/10.1007/JHEP02(2016)107}{\emph{JHEP} {\bf 02} (2016)
  107}, [\href{http://arxiv.org/abs/1512.01963}{{\tt 1512.01963}}].

\bibitem{Bernal:2012gv}
N.~Bernal, F.-X. Josse-Michaux and L.~Ubaldi, \emph{{Phenomenology of WIMPy
  baryogenesis models}},
  \href{http://dx.doi.org/10.1088/1475-7516/2013/01/034}{\emph{JCAP} {\bf 1301}
  (2013) 034}, [\href{http://arxiv.org/abs/1210.0094}{{\tt 1210.0094}}].

\bibitem{Zhang:2012cd}
C.~Zhang, N.~Greiner and S.~Willenbrock, \emph{{Constraints on Non-standard Top
  Quark Couplings}},
  \href{http://dx.doi.org/10.1103/PhysRevD.86.014024}{\emph{Phys. Rev.} {\bf
  D86} (2012) 014024}, [\href{http://arxiv.org/abs/1201.6670}{{\tt
  1201.6670}}].

\bibitem{Herrero-Garcia:2016uab}
J.~Herrero-Garcia, N.~Rius and A.~Santamaria, \emph{{Higgs lepton flavour
  violation: UV completions and connection to neutrino masses}},
  \href{http://arxiv.org/abs/1605.06091}{{\tt 1605.06091}}.

\bibitem{Yang:1997iv}
J.~M. Yang and B.-L. Young, \emph{{Dimension-six CP violating operators of the
  third family quarks and their effects at colliders}},
  \href{http://dx.doi.org/10.1103/PhysRevD.56.5907}{\emph{Phys. Rev.} {\bf D56}
  (1997) 5907--5918}, [\href{http://arxiv.org/abs/hep-ph/9703463}{{\tt
  hep-ph/9703463}}].

\bibitem{Whisnant:1997qu}
K.~Whisnant, J.-M. Yang, B.-L. Young and X.~Zhang, \emph{{Dimension-six CP
  conserving operators of the third family quarks and their effects on collider
  observables}}, \href{http://dx.doi.org/10.1103/PhysRevD.56.467}{\emph{Phys.
  Rev.} {\bf D56} (1997) 467--478},
  [\href{http://arxiv.org/abs/hep-ph/9702305}{{\tt hep-ph/9702305}}].

\bibitem{AguilarSaavedra:2009mx}
J.~A. Aguilar-Saavedra, \emph{{A Minimal set of top-Higgs anomalous
  couplings}},
  \href{http://dx.doi.org/10.1016/j.nuclphysb.2009.06.022}{\emph{Nucl. Phys.}
  {\bf B821} (2009) 215--227}, [\href{http://arxiv.org/abs/0904.2387}{{\tt
  0904.2387}}].

\bibitem{Sakharov:1967dj}
A.~D. Sakharov, \emph{{Violation of $CP$ invariance, $c$ asymmetry, and baryon
  asymmetry of the universe}},
  \href{http://dx.doi.org/10.1070/PU1991v034n05ABEH002497}{\emph{Pis'ma Zh.
  Eksp. Teor. Fiz.} {\bf 5} (1967) 32--35}.

\bibitem{Nomura:2009tw}
D.~Nomura, \emph{{Effects of Top-quark Compositeness on Higgs Boson Production
  at the LHC}}, \href{http://dx.doi.org/10.1007/JHEP02(2010)061}{\emph{JHEP}
  {\bf 02} (2010) 061}, [\href{http://arxiv.org/abs/0911.1941}{{\tt
  0911.1941}}].

\bibitem{Grzadkowski:2010es}
B.~Grzadkowski, M.~Iskrzynski, M.~Misiak and J.~Rosiek, \emph{{Dimension-Six
  Terms in the Standard Model Lagrangian}},
  \href{http://dx.doi.org/10.1007/JHEP10(2010)085}{\emph{JHEP} {\bf 10} (2010)
  085}, [\href{http://arxiv.org/abs/1008.4884}{{\tt 1008.4884}}].

\bibitem{Martin:1959jp}
P.~C. Martin and J.~Schwinger, \emph{{Theory of Many Particle Systems. I}},
  \href{http://dx.doi.org/10.1103/PhysRev.115.1342}{\emph{Phys. Rev.} {\bf 115}
  (1959) 1342--1373}.

\bibitem{Schwinger:1960qe}
J.~S. Schwinger, \emph{{Brownian Motion of a Quantum Oscillator}},
  \href{http://dx.doi.org/10.1063/1.1703727}{\emph{J. Math. Phys.} {\bf 2}
  (1961) 407--432}.

\bibitem{Keldysh:1964ud}
L.~V. Keldysh, \emph{{Diagram technique for nonequilibrium processes}},
  {\emph{Zh. Eksp. Teor. Fiz.} {\bf 47} (1964) 1515--1527}.

\bibitem{Chou:1984es}
K.-c. Chou, Z.-b. Su, B.-l. Hao and L.~Yu, \emph{{Equilibrium and
  nonequilibrium formalisms made unified}},
  \href{http://dx.doi.org/10.1016/0370-1573(85)90136-X}{\emph{Phys. Rept.} {\bf
  118} (1985) 1}.

\bibitem{Mahanthappa:1962ex}
K.~T. Mahanthappa, \emph{{Multiple Production of Photons in Quantum
  Electrodynamics}},
  \href{http://dx.doi.org/10.1103/PhysRev.126.329}{\emph{Phys. Rev.} {\bf 126}
  (1962) 329--340}.

\bibitem{Cirigliano:2009yt}
V.~Cirigliano, C.~Lee, M.~J. Ramsey-Musolf and S.~Tulin, \emph{{Flavored
  Quantum Boltzmann Equations}},
  \href{http://dx.doi.org/10.1103/PhysRevD.81.103503}{\emph{Phys. Rev.} {\bf
  D81} (2010) 103503}, [\href{http://arxiv.org/abs/0912.3523}{{\tt
  0912.3523}}].

\bibitem{Cirigliano:2011di}
V.~Cirigliano, C.~Lee and S.~Tulin, \emph{{Resonant Flavor Oscillations in
  Electroweak Baryogenesis}},
  \href{http://dx.doi.org/10.1103/PhysRevD.84.056006}{\emph{Phys. Rev.} {\bf
  D84} (2011) 056006}, [\href{http://arxiv.org/abs/1106.0747}{{\tt
  1106.0747}}].

\bibitem{Weldon:1989ys}
H.~A. Weldon, \emph{{Dynamical Holes in the Quark - Gluon Plasma}},
  \href{http://dx.doi.org/10.1103/PhysRevD.40.2410}{\emph{Phys. Rev.} {\bf D40}
  (1989) 2410}.

\bibitem{Weldon:1999th}
H.~A. Weldon, \emph{{Structure of the quark propagator at high temperature}},
  \href{http://dx.doi.org/10.1103/PhysRevD.61.036003}{\emph{Phys. Rev.} {\bf
  D61} (2000) 036003}, [\href{http://arxiv.org/abs/hep-ph/9908204}{{\tt
  hep-ph/9908204}}].

\bibitem{Klimov:1981ka}
V.~V. Klimov, \emph{{Spectrum of Elementary Fermi Excitations in Quark Gluon
  Plasma. (In Russian)}}, {\emph{Sov. J. Nucl. Phys.} {\bf 33} (1981)
  934--935}.

\bibitem{Lee:2004we}
C.~Lee, V.~Cirigliano and M.~J. Ramsey-Musolf, \emph{{Resonant relaxation in
  electroweak baryogenesis}},
  \href{http://dx.doi.org/10.1103/PhysRevD.71.075010}{\emph{Phys. Rev. D} {\bf
  71} (2005) 075010}, [\href{http://arxiv.org/abs/hep-ph/0412354}{{\tt
  hep-ph/0412354}}].

\bibitem{Cirigliano:2006wh}
V.~Cirigliano, M.~J. Ramsey-Musolf, S.~Tulin and C.~Lee, \emph{{Yukawa and
  triscalar processes in electroweak baryogenesis}},
  \href{http://dx.doi.org/10.1103/PhysRevD.73.115009}{\emph{Phys. Rev. D} {\bf
  73} (2006) 115009}, [\href{http://arxiv.org/abs/hep-ph/0603058}{{\tt
  hep-ph/0603058}}].

\bibitem{White:2015bva}
G.~A. White, \emph{{General analytic methods for solving coupled transport
  equations: From cosmology to beyond}},
  \href{http://dx.doi.org/10.1103/PhysRevD.93.043504}{\emph{Phys. Rev.} {\bf
  D93} (2016) 043504}, [\href{http://arxiv.org/abs/1510.03901}{{\tt
  1510.03901}}].

\bibitem{Carena:2002ss}
M.~Carena, M.~Quir\'os, M.~Seco and C.~E.~M. Wagner, \emph{{Improved results in
  supersymmetric electroweak baryogenesis}},
  \href{http://dx.doi.org/10.1016/S0550-3213(02)01065-9}{\emph{Nucl. Phys. B}
  {\bf 650} (2003) 24--42}, [\href{http://arxiv.org/abs/hep-ph/0208043}{{\tt
  hep-ph/0208043}}].

\bibitem{Cline:2000nw}
J.~M. Cline, M.~Joyce and K.~Kainulainen, \emph{{Supersymmetric electroweak
  baryogenesis}},
  \href{http://dx.doi.org/10.1088/1126-6708/2000/07/018}{\emph{JHEP} {\bf 07}
  (2000) 018}, [\href{http://arxiv.org/abs/hep-ph/0006119}{{\tt
  hep-ph/0006119}}].

\bibitem{Bodeker:1999gx}
D.~B{\"o}deker, G.~D. Moore and K.~Rummukainen, \emph{{Chern-Simons number
  diffusion and hard thermal loops on the lattice}},
  \href{http://dx.doi.org/10.1103/PhysRevD.61.056003}{\emph{Phys. Rev. D} {\bf
  61} (2000) 056003}, [\href{http://arxiv.org/abs/hep-ph/9907545}{{\tt
  hep-ph/9907545}}].

\bibitem{Moore:1999fs}
G.~D. Moore and K.~Rummukainen, \emph{{Classical sphaleron rate on fine
  lattices}}, \href{http://dx.doi.org/10.1103/PhysRevD.61.105008}{\emph{Phys.
  Rev. D} {\bf 61} (2000) 105008},
  [\href{http://arxiv.org/abs/hep-ph/9906259}{{\tt hep-ph/9906259}}].

\bibitem{Moore:2000mx}
G.~D. Moore, \emph{{Sphaleron rate in the symmetric electroweak phase}},
  \href{http://dx.doi.org/10.1103/PhysRevD.62.085011}{\emph{Phys. Rev. D} {\bf
  62} (2000) 085011}, [\href{http://arxiv.org/abs/hep-ph/0001216}{{\tt
  hep-ph/0001216}}].

\bibitem{Baron:2013eja}
{\scshape ACME} collaboration, J.~Baron et~al., \emph{{Order of Magnitude
  Smaller Limit on the Electric Dipole Moment of the Electron}},
  \href{http://dx.doi.org/10.1126/science.1248213}{\emph{Science} {\bf 343}
  (2014) 269--272}, [\href{http://arxiv.org/abs/1310.7534}{{\tt 1310.7534}}].

\bibitem{Bian:2014zka}
L.~Bian, T.~Liu and J.~Shu, \emph{{Cancellations Between Two-Loop Contributions
  to the Electron Electric Dipole Moment with a CP-Violating Higgs Sector}},
  \href{http://dx.doi.org/10.1103/PhysRevLett.115.021801}{\emph{Phys. Rev.
  Lett.} {\bf 115} (2015) 021801}, [\href{http://arxiv.org/abs/1411.6695}{{\tt
  1411.6695}}].

\bibitem{Crivellin:2013hpa}
A.~Crivellin, S.~Najjari and J.~Rosiek, \emph{{Lepton Flavor Violation in the
  Standard Model with general Dimension-Six Operators}},
  \href{http://dx.doi.org/10.1007/JHEP04(2014)167}{\emph{JHEP} {\bf 04} (2014)
  167}, [\href{http://arxiv.org/abs/1312.0634}{{\tt 1312.0634}}].

\bibitem{Regan:2002ta}
B.~C. Regan, E.~D. Commins, C.~J. Schmidt and D.~DeMille, \emph{{New limit on
  the electron electric dipole moment}},
  \href{http://dx.doi.org/10.1103/PhysRevLett.88.071805}{\emph{Phys. Rev.
  Lett.} {\bf 88} (2002) 071805}.

\bibitem{Griffith:2009zz}
W.~C. Griffith, M.~D. Swallows, T.~H. Loftus, M.~V. Romalis, B.~R. Heckel and
  E.~N. Fortson, \emph{{Improved Limit on the Permanent Electric Dipole Moment
  of Hg-199}},
  \href{http://dx.doi.org/10.1103/PhysRevLett.102.101601}{\emph{Phys. Rev.
  Lett.} {\bf 102} (2009) 101601}.

\bibitem{Afach:2015sja}
J.~M. Pendlebury et~al., \emph{{Revised experimental upper limit on the
  electric dipole moment of the neutron}},
  \href{http://dx.doi.org/10.1103/PhysRevD.92.092003}{\emph{Phys. Rev.} {\bf
  D92} (2015) 092003}, [\href{http://arxiv.org/abs/1509.04411}{{\tt
  1509.04411}}].

\bibitem{Baker:2006ts}
C.~A. Baker et~al., \emph{{An Improved experimental limit on the electric
  dipole moment of the neutron}},
  \href{http://dx.doi.org/10.1103/PhysRevLett.97.131801}{\emph{Phys. Rev.
  Lett.} {\bf 97} (2006) 131801},
  [\href{http://arxiv.org/abs/hep-ex/0602020}{{\tt hep-ex/0602020}}].

\bibitem{Barr:1990vd}
S.~M. Barr and A.~Zee, \emph{{Electric Dipole Moment of the Electron and of the
  Neutron}}, \href{http://dx.doi.org/10.1103/PhysRevLett.65.21}{\emph{Phys.
  Rev. Lett.} {\bf 65} (1990) 21--24}.

\bibitem{Yamanaka:2014mda}
N.~Yamanaka, \emph{{Analysis of the Electric Dipole Moment in the R-parity
  Violating Supersymmetric Standard Model}}.
\newblock PhD thesis, Osaka U., Res. Ctr. Nucl. Phys., 2013.
\newblock 10.1007/978-4-431-54544-6.

\bibitem{Yamanaka:2012hm}
N.~Yamanaka, T.~Sato and T.~Kubota, \emph{{A Reappraisal of two-loop
  contributions to the fermion electric dipole moments in R-parity violating
  supersymmetric models}},
  \href{http://dx.doi.org/10.1103/PhysRevD.85.117701}{\emph{Phys. Rev.} {\bf
  D85} (2012) 117701}, [\href{http://arxiv.org/abs/1202.0106}{{\tt
  1202.0106}}].

\bibitem{Stockinger:2006zn}
D.~Stockinger, \emph{{The Muon Magnetic Moment and Supersymmetry}},
  \href{http://dx.doi.org/10.1088/0954-3899/34/2/R01}{\emph{J. Phys.} {\bf G34}
  (2007) R45--R92}, [\href{http://arxiv.org/abs/hep-ph/0609168}{{\tt
  hep-ph/0609168}}].

\bibitem{Leigh:1990kf}
R.~G. Leigh, S.~Paban and R.~M. Xu, \emph{{Electric dipole moment of
  electron}}, \href{http://dx.doi.org/10.1016/0550-3213(91)90128-K}{\emph{Nucl.
  Phys.} {\bf B352} (1991) 45--58}.

\bibitem{Abe:2013qla}
T.~Abe, J.~Hisano, T.~Kitahara and K.~Tobioka, \emph{{Gauge invariant Barr-Zee
  type contributions to fermionic EDMs in the two-Higgs doublet models}},
  \href{http://dx.doi.org/10.1007/JHEP01(2014)106,
  10.1007/JHEP04(2016)161}{\emph{JHEP} {\bf 01} (2014) 106},
  [\href{http://arxiv.org/abs/1311.4704}{{\tt 1311.4704}}].

\bibitem{Jung:2013hka}
M.~Jung and A.~Pich, \emph{{Electric Dipole Moments in Two-Higgs-Doublet
  Models}}, \href{http://dx.doi.org/10.1007/JHEP04(2014)076}{\emph{JHEP} {\bf
  04} (2014) 076}, [\href{http://arxiv.org/abs/1308.6283}{{\tt 1308.6283}}].

\bibitem{BowserChao:1997bb}
D.~Bowser-Chao, D.~Chang and W.-Y. Keung, \emph{{Electron electric dipole
  moment from CP violation in the charged Higgs sector}},
  \href{http://dx.doi.org/10.1103/PhysRevLett.79.1988}{\emph{Phys. Rev. Lett.}
  {\bf 79} (1997) 1988--1991}, [\href{http://arxiv.org/abs/hep-ph/9703435}{{\tt
  hep-ph/9703435}}].

\bibitem{Altmannshofer:2015qra}
W.~Altmannshofer, J.~Brod and M.~Schmaltz, \emph{{Experimental constraints on
  the coupling of the Higgs boson to electrons}},
  \href{http://dx.doi.org/10.1007/JHEP05(2015)125}{\emph{JHEP} {\bf 05} (2015)
  125}, [\href{http://arxiv.org/abs/1503.04830}{{\tt 1503.04830}}].

\bibitem{Profumo:2007wc}
S.~Profumo, M.~J. Ramsey-Musolf and G.~Shaughnessy, \emph{{Singlet Higgs
  phenomenology and the electroweak phase transition}},
  \href{http://dx.doi.org/10.1088/1126-6708/2007/08/010}{\emph{JHEP} {\bf 08}
  (2007) 010}, [\href{http://arxiv.org/abs/0705.2425}{{\tt 0705.2425}}].

\bibitem{Patel:2011th}
H.~H. Patel and M.~J. Ramsey-Musolf, \emph{{Baryon Washout, Electroweak Phase
  Transition, and Perturbation Theory}},
  \href{http://dx.doi.org/10.1007/JHEP07(2011)029}{\emph{JHEP} {\bf 07} (2011)
  029}, [\href{http://arxiv.org/abs/1101.4665}{{\tt 1101.4665}}].

\bibitem{Patel:2012pi}
H.~H. Patel and M.~J. Ramsey-Musolf, \emph{{Stepping Into Electroweak Symmetry
  Breaking: Phase Transitions and Higgs Phenomenology}},
  \href{http://dx.doi.org/10.1103/PhysRevD.88.035013}{\emph{Phys. Rev.} {\bf
  D88} (2013) 035013}, [\href{http://arxiv.org/abs/1212.5652}{{\tt
  1212.5652}}].

\bibitem{Tulin:2011wi}
S.~Tulin and P.~Winslow, \emph{{Anomalous $B$ meson mixing and baryogenesis}},
  \href{http://dx.doi.org/10.1103/PhysRevD.84.034013}{\emph{Phys. Rev.} {\bf
  D84} (2011) 034013}, [\href{http://arxiv.org/abs/1105.2848}{{\tt
  1105.2848}}].

\end{thebibliography}\endgroup

\end{document}